\documentclass[]{aa}  

\usepackage{graphicx}
\usepackage{txfonts}
\usepackage{graphicx}   
\usepackage{amsmath}    

\usepackage{threeparttable}
\usepackage{tabularx}
\usepackage{booktabs} 

\usepackage[dvipsnames]{xcolor}
\usepackage{rotating} 

\usepackage{multicol}
\usepackage{multirow}

\usepackage{natbib}
\usepackage{placeins}
\usepackage[colorlinks=true, linkcolor=blue, citecolor=blue]{hyperref}
\usepackage{amssymb}

\newcommand{\ci}{[C {\sc i}]}
\newcommand{\cii}{[C {\sc ii}]}
\newcommand{\oiii}{[O\,{\footnotesize III}]}

\newcommand{\nii}{[N\,{\footnotesize II}]}

\newcommand{\hii}{H\,{\footnotesize II}}

\newcommand{\cofour}{CO\,$(4-3)$}

\newcommand{\jwst}{JWST}
\newcommand{\hst}{HST}
\newcommand{\spitzer}{\textit{Spitzer}}

\newcommand\tb[1]{\textbf{#1}}

\definecolor{purple}{rgb}{0.6, 0.4, 0.8}

\newcommand{\rdate}{2025-03-26}
\newcommand{\surveyarea}{391}
\newcommand{\totalarea}{916}
\newcommand{\nimage}{1847}
\newcommand{\sncutblind}{6.6}
\newcommand{\sncutprior}{5.1}

\newcommand{\NalldetpriorRes}{1262}
\newcommand{\NalldetpriorResUniq}{614}
\newcommand{\NalldetpriorResUniqz}{307}

\newcommand{\Nalldetblind}{1070}
\newcommand{\Nalldetblindopt}{858}
\newcommand{\Nuniqdetblindopt}{408}
\newcommand{\Nblindnocounterpart}{143}
\newcommand{\Nblindnocounterpartuniq}{96}

\newcommand{\NSecure}{832}

\newcommand{\NuniqSecure}{395}

\newcommand{\Nall}{1288}
\newcommand{\Nallz}{672}
\newcommand{\Nalluniq}{622}
\newcommand{\Nalluniqz}{307}

\newcommand{\NgaltotALMA}{128,125}
\newcommand{\NzspecALMA}{11,537}
\newcommand{\NzDJAALMA}{3880}
\newcommand{\Ntotparent}{334,811}
\newcommand{\Nzspec}{19,236}
\newcommand{\NzspecSecure}{16,874}
\newcommand{\Nzspecoverlap}{4137}
\newcommand{\NzspecHST}{7761}

\newcommand{\NzspecALL}{24,635}

\newcommand{\djaspecversion}{v4.3}
\newcommand{\djaphotversion}{v7.0}

\newcommand{\acosmos}{A$^{3}$COSMOS}
\newcommand{\agds}{A$^{3}$GOODSS}
\newcommand{\ecogal}{ECOGAL}
\newcommand{\eazypy}{\textsc{eazypy}}

\begin{document} 

   \title{ECOGAL}

   \subtitle{I. Project design and the first catalogue}

   \author{Minju Lee\inst{1,2}
   \and Georgios Magdis\inst{1,2,3}
   \and Gabriel Brammer\inst{1,3}
   \and Daizhong Liu\inst{4}
   \and Benjamin Magnelli\inst{5}
   \and Steven Gillman\inst{1,2}
   \and Bitten Gullberg\inst{1,2}
   \and Kei Ito\inst{1,2}
   \and Nikolaj Sillassen\inst{1,2}
   \and Francesco Valentino\inst{1,2}
   \and Rashmi Gottumukkala\inst{1,3}
   \and Chandana Chandrashekhar\inst{1,2}
   \and Thomas R. Greve\inst{1,2}
          }

\institute{Cosmic Dawn Center (DAWN), Copenhagen, Denmark
\and 
DTU Space, Technical University of Denmark, Elektrovej 327, DK2800 Kgs. Lyngby, Denmark
\and
Niels Bohr Institute, University of Copenhagen, Jagtvej 128, 2200 Copenhagen N, Denmark
\and
Purple Mountain Observatory, Chinese Academy of Sciences, 10
Yuanhua Road, Nanjing 210023, China
\and
Universite Paris-Saclay, Universite Paris Cite, CEA, CNRS, AIM,
91191 Gif-sur-Yvette, France
}
   \date{Received; accepted}
  \abstract
   {We present ECOlogy for Galaxies using ALMA archive and Legacy surveys (\ecogal), an ALMA data mining project. Using the footprints of the James Webb Space Telescope (JWST) and the \textit{Hubble} Space Telescope (HST), we query and uniformly reprocess ALMA data to produce continuum images and two complementary source catalogues: (i) a prior-based catalogue anchored to optical/near-infrared detections, and (ii) a blind catalogue based on significant ALMA detections. Detection thresholds are established through peak signal-to-noise-ratio statistics and analysis of inverted maps. In this paper, we focus on the ALMA-accessible CANDELS fields (COSMOS, GOODS-S, and UDS), covering $\sim$130,000 optical/NIR-selected galaxies spanning $0.0<z<15.0$. We identify \Nall\ detections (\Nalluniq\ unique) with optical/NIR counterparts across the two methods, of which
   \NSecure\ detections (\NuniqSecure\ unique) appear in both.
   Among the \Nalluniq\ unique sources, \Nalluniqz\ have spectroscopic redshifts from $0.12<z<6.85$. 
   \ecogal\ expands the parameter space previously explored by \acosmos\, and A$^3$GOODS-S by incorporating the UDS field and integrating publicly available JWST datasets from the DAWN JWST Archive (DJA). We also highlight several science cases enabled by \ecogal, including studies of gas fractions inferred from dust continuum emission in galaxies across different environments, the obscured fraction of star formation, and optically dark systems. \ecogal\ provides science-ready datasets for the community and showcases the power of combining over a decade of accumulated public data with other legacy datasets. The data product is made publicly available and accompanies a post demonstrating the usage of the catalogue on the DJA webpage.}


   \maketitle
   \nolinenumbers


\section{Introduction}
The formation and evolution of galaxies are among the most complex processes in our understanding of the Universe’s history.
A comprehensive understanding of galaxy formation and evolution requires observations spanning the entire electromagnetic spectrum.
Over the past three decades, an avalanche of datasets from rest-frame optical and near-infrared (NIR) instruments has allowed us to study the cosmic star formation history and the following stellar mass build-up of galaxies (e.g. \citealt{Madau2014, Traina2024, Magnelli2024, Shuntov2025}) and possible connections with the growth of the black holes (e.g. \citealt{Kormendy2013, Heckman2014}) over cosmic time.
Equally transformative is the advent of the Atacama Large Millimeter/submillimeter Array (ALMA), whose submillimetre/millimetre observations open a complementary window onto galaxies’ cold interstellar medium. 
Since 2011, ALMA has delivered measurements of molecular gas, dust, and obscured star formation at an unprecedented sensitivity and angular resolution, providing key constraints on the fuel for star formation and the physical processes that regulate galaxy evolution (e.g. \citealt{Scoville2014, Scoville2016, Walter2020, Wang2022}; see also the review by \citealt{Tacconi2020}).
ALMA’s long-baseline configurations also achieve angular resolutions comparable to -- or better than -- 8-10 m class ground-based optical/NIR facilities, enabling spatially resolved kinematic studies of high-redshift galaxies using cold-gas tracers (e.g. \citealt{deBreuck2014, minju2019a, Rizzo2020, Genzel2023, Neeleman2017, Meyer2025, LeeL2025, NestorShachar2025}).

After more than a decade of operation, ALMA has accumulated sufficient sky coverage to enable systematic studies of the cosmic evolution of gas and dust content (e.g. \citealt{Liu2019a, Adscheid2024}).
\acosmos\footnote{\url{https://sites.google.com/view/a3cosmos}} and \agds\ projects \citep{Liu2019a, Liu2019b, Adscheid2024} 
demonstrated the scientific potential of the ALMA archive by identifying over 2000 unique sources in the Cosmic Evolution Survey (COSMOS; \citealt{Scoville2007}) and the Great Observatories Origins Deep Survey–South (GOODS-S; \citealt{Giavalisco2004}) legacy fields.

Building upon this foundation, \tb{ECO}logy for \tb{G}alaxies using \tb{A}LMA archive and \tb{L}egacy surveys (hereafter \ecogal) is an independent effort focusing on fields observed with the \textit{James Webb} Space Telescope (\jwst) and the \textit{Hubble} Space Telescope (\hst).
Analogous to \acosmos\, and \agds, \ecogal\ mines and processes ALMA archival datasets to fully leverage the legacy value of ALMA, JWST, and HST.

We first aim to collect regions in three major legacy fields targeted by Cosmic Assembly Near-infrared Deep Extragalactic Legacy Survey (CANDELS) and 3D-\hst\ \citep{Grogin2011, Koekemoer2011, Brammer2012} that are observable from the ALMA site. 
The CANDELS field covers a smaller part of the entire COSMOS and GOODS-S areas, both of which have already been processed \acosmos. 
However, \ecogal\ (i) combines publicly available spectroscopic redshift information from \jwst\ or elsewhere, building a panchromatic database for the scientific community, and (ii) adds the UKIDSS Ultra-Deep Survey (UDS; \citealt{Lawrence2007}) field, intending to maximise the accessibility and enhance the legacy values of \ecogal.
The first catalogue and the image product focus on the dust continuum properties. 

This paper is structured as follows. 
In Sect.~\ref{sec:parent}, we describe the parent catalogue constructed from available photometry and spectroscopy from the DAWN JWST Archive (DJA; \citealt{grizli, msaexp, Valentino2023, deGraaff2025a, Heintz2024}), the 3D-HST catalogue \citep{Brammer2012}, and the literature to search for ALMA data. 
Section~\ref{sec:dataprocessing} explains the core part of the ALMA data mining effort; namely, the data retrieval, reduction, and imaging. 
The section also shows how we build the catalogues based on the prior positions and blind detection, and assess the detections. 
Section~\ref{sec:galaxyproperties} presents the properties of \ecogal\, galaxies in terms of their properties, such as rest-frame $UVJ$ colours, stellar masses ($M_{\star}$), and star formation rates (SFRs).
We provide several science cases that can be explored by \ecogal\, data products in Sect.~\ref{sec:science} and a summary in Sect.~\ref{sec:summary}.

In the following, we assume a flat $\Lambda$ cold dark matter cosmology with $H_0 = 70$ km s$^{-1}$ Mpc$^{-1}$, $\Omega_{\Lambda} = 0.7$, $\Omega_{M} = 0.3$, and a Chabrier initial mass function \citep{Chabrier2003}. All magnitudes are expressed in the AB system \citep{Oke1983}.


\section{Parent catalogue construction}\label{sec:parent}

\subsection{Prior photometric catalogue}\label{sec:catalog}

We constructed the parent sample from two publicly available catalogues: the 3D-\hst\,\footnote{\url{https://archive.stsci.edu/prepds/3d-hst/}} and DJA \footnote{\url{https://dawn-cph.github.io/dja/index.html}}.
The 3D-\hst\ survey covers $\sim$600 square-arcmin across well-studied extragalactic fields (AEGIS, COSMOS, GOODS-S/N, and UKIDSS-UDS) and builds on the CANDELS survey 
\citep{Grogin2011, Koekemoer2011}. 
It provides grism redshifts for $\approx$7,000 galaxies at $1<z<3.5$ \citep{Brammer2012}, greatly expanding the sample of spectroscopically confirmed sources beyond $z\sim$1.
These data have enabled a broad range of studies on galaxy formation and evolution, including stellar mass assembly, structural transformation, quenching, and environmental effects (e.g. \citealt{vanDokkum2013, Wuyts2013, vanderwel2014, Hill2019, Belli2019, Fossati2017}).

Since the launch, \jwst\, has extensively observed these same fields, revealing redder \citep{deGraaff2025a}, more distant \citep{Carniani2024, Zavala2025, Curti2025}, and otherwise unique galaxy populations \citep{Matthee2024, Perez-Gonzalez2024, Isobe2023}, and providing rest-frame optical coverage that improves constraints on stellar populations and morphology (e.g. \citealt{Slob2024, Shapley2025, EspejoSalcedo2025}).
DJA provides publicly reduced \jwst\ imaging and spectroscopy data using \texttt{grizli}\footnote{\url{https://github.com/gbrammer/grizli}} and \texttt{msaexp}\footnote{\url{https://github.com/gbrammer/msaexp}}, delivering homogeneous photometric (\djaphotversion\, or later; \citealt{Valentino2023}) and spectroscopic data products (\djaspecversion; \citealt{Heintz2024, deGraaff2025a, Ito2026}).
We cross-matched the 3D-HST and DJA catalogues after applying astrometric corrections (Sect.~\ref{sec:astrometry}), producing a unified parent sample in the COSMOS, GOODS-S, and UDS fields that includes a total of \Ntotparent\,extragalactic sources.
Based on this parent catalogue, we queried all galaxies whose positions fall within the ALMA observational footprints (Sect.~\ref{sec:dataprocessing}). Figure~\ref{fig:coverage} illustrates an example in GOODS-S, showing that the ALMA footprints (shaded grey regions) do not provide contiguous coverage of the JWST/HST survey area. Because the ALMA archive consists primarily of individual pointed observations rather than contiguous survey fields, only \NgaltotALMA\ galaxies from the parent catalogue are included in the first ECOGAL data release.

\subsection{Astrometry correction in the 3D-HST catalogue}\label{sec:astrometry}
For the CANDELS/3D-HST catalogue, previous studies have reported systematic offsets relative to ALMA in GOODS-S \citep{Dunlop2017} and COSMOS \citep{Liu2019a}.
Accordingly, we applied median offsets to the 3D-HST positions relative to the DJA JWST astrometry, which was corrected based on the \textit{Gaia} position, when constructing the parent catalogue in combination with DJA.

The corrections are negligible in COSMOS and UDS ($<0\farcs1$ in RA and Dec), but larger offsets are required in GOODS-S ($-0\farcs101$ in RA, $+0\farcs262$ in Dec), consistent with \citet{Franco2018}.
Figure~\ref{fig:astrometry} shows the distribution of positional offsets between \jwst\ and 3D-\hst.
After the median correction, a residual scatter of $\approx0\farcs1$ remains, primarily due to local effects such as dust attenuation, internal structural differences, and mosaic distortions \citep{Gomez-Guijarro2022}.
Given that the median (mean) ALMA resolution is $0\farcs70$ ($0\farcs87$), we do not correct for these small-scale distortions.
We discuss the astrometry compared to ALMA in more detail in Sect.~\ref{sec:almaastrometry}.

\subsection{Redshift information}
\subsubsection{Spectroscopic redshift information}\label{sec:zcatalog}
We compiled spectroscopic redshift information for the parent catalogue by combining publicly available datasets from multiple surveys:
C3R2 \citep[DR2+3]{Masters2019, Stanford2021},
CANDELSz7 \citep{Pentericci2018},
FMOS \citep[v2]{Silverman2015, Kashino2019},
GAMA \citep[DR4]{Driver2022},
GOLDRUSH \citep{Harikane2022},
DEIMOS-COSMOS \citep{Hasinger2018},
JADES \citep{Bunker2024},
KMOS3D \citep{Wisnioski2019},
LEGA-C \citep[DR3]{vanderwel2021},
MOSDEF \citep{Kriek2015},
MUSE WIDE \citep[DR1]{Urrutia2019},
PRIMUS \citep[DR1]{Coil2011,Cool2013},
UDSz \citep{Bradshaw2013,McLure2013},
VANDELS \citep[DR4]{Garilli2021},
VIPERS \citep[PDR-2]{Scodeggio2018},
VUDS \citep[DR1]{Tasca2017},
VVDS \citep{LeFevre2013},
zFIRE \citep[DR1]{Nanayakkara2016},
zCOSMOS \citep[DR3]{Lilly2009}, the latest COSMOS spectroscopic catalogue \citep{Khostovan2026}\footnote{\url{https://github.com/cosmosastro/speczcompilation}}, and 
the DJA spectra \citep[v4.3]{Valentino2025, Ito2026}\footnote{\url{https://dawn-cph.github.io/dja/blog/2025/05/01/nirspec-merged-table-v4/}}.

\begin{figure*}[th]
\sidecaption
\includegraphics[width=12cm]{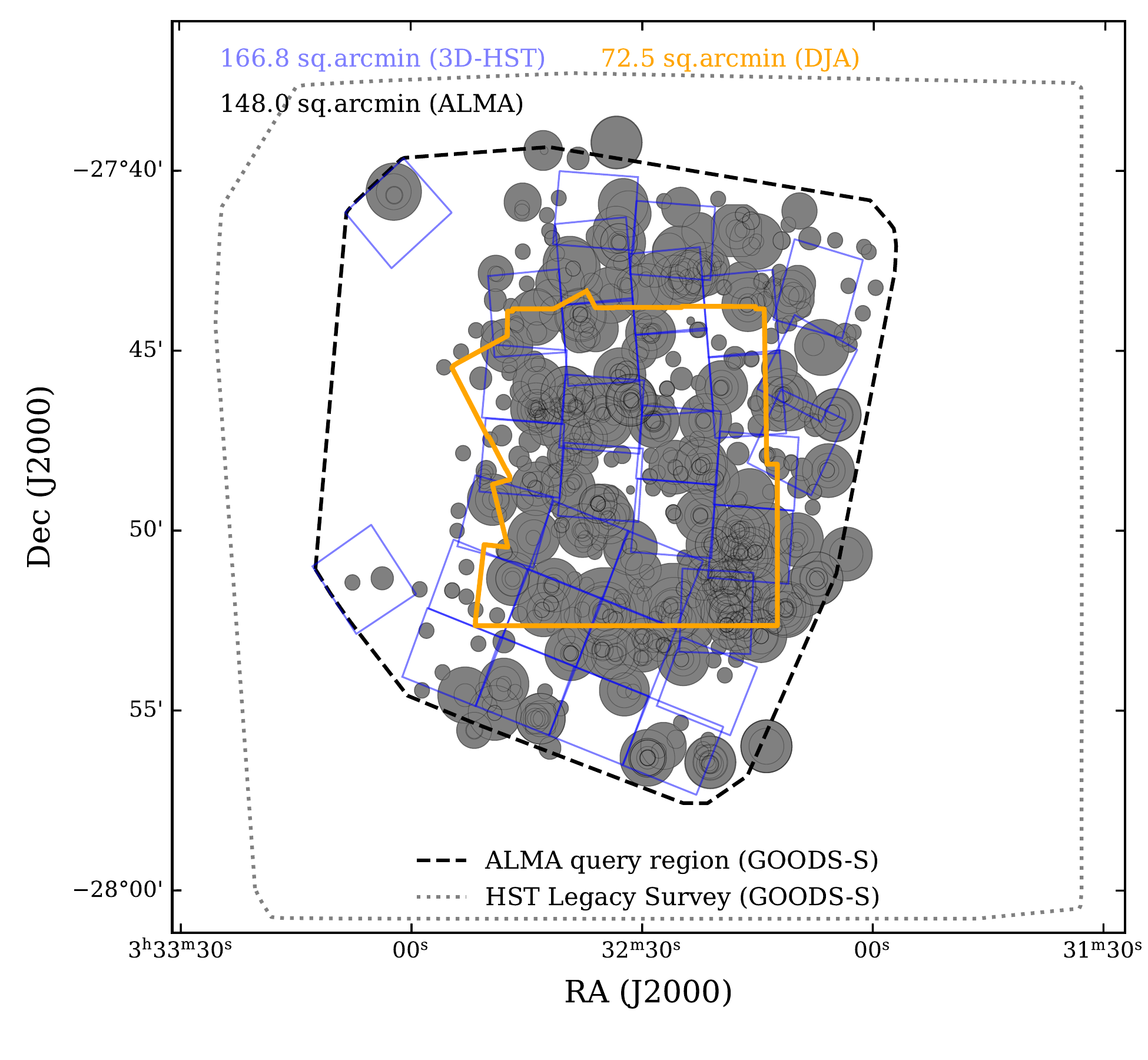}
\caption{ALMA observations in the GOODS-S region used for the first data release (shown as filled grey polygons). Only single-pointing observations are included. Concentric circles indicate multiple band and/or depth observations for the same target.
The dotted region indicates the Hubble Legacy
Fields \citep{Illingworth2016, Whitaker2019} to indicate the extent of the GOODS-S field. The blue rectangle outlines the 3D-\hst\ F140W coverage, while the orange polygon marks the \jwst\ F444W mosaic footprint from DJA. The dashed line shows the ALMA archive query region, chosen to encompass the full 3D-HST and JWST coverage.\label{fig:coverage}}
\end{figure*}

We cross-matched the sources to the parent catalogue using a matching radius of $0\farcs5$.
From the \Ntotparent\ galaxies in the parent catalogue, we retrieved 
\Nzspec\ spectroscopic redshifts, of which \NzspecSecure\ are classified as robust (Appendix~\ref{app:specz}).
We additionally include spectroscopic redshifts from the 3D-HST catalogue, which includes both slitless grism redshifts \citep{Momcheva2016} and ground-based spectroscopic measurements \citep{Skelton2014}; Column \texttt{z\_best\_s} = 1 and 2 in the 3D-HST catalogue\footnote{In the 3D-HST data release, the grism redshifts were obtained through a simultaneous fit to the slitless grism spectra and multi-band photometry, with a subset of sources ($JH_{\rm IR}<24$) additionally inspected visually.}.
Of the secure 3D-HST redshifts, \Nzspecoverlap\ overlap with our compilation, while \NzspecHST\ are unique. Incorporating these additional measurements increases the number of secure spectroscopic entries in the parent catalogue to \NzspecALL\ ($\approx7\%$ of the parent sample).

\subsubsection{Photometric redshift estimation}\label{sec:photoz}
We estimated photometric redshifts using \eazypy\ \citep{Brammer2008, Brammer2021}\footnote{\url{https://github.com/gbrammer/eazy-py}}, using the multi-wavelength photometry compiled in the parent catalogue. This combines the original 3D-HST photometry with the JWST/DJA measurements.
In addition to the HST bands, the 3D-HST catalogue includes ground-based optical and NIR photometry together with \spitzer\ imaging \citep{Skelton2014}. Altogether, each field has more than 30 photometric bands. The filters used in the \eazypy \,fitting are summarised in Table~\ref{tab:filters}.

\begin{table}[]
\caption{Photometric filters and the corresponding instrument compiled in the parent catalogue.}
\small
\begin{tabular}{l|l|l}
\hline \hline
Field   & Filters                                                                        & Telescope/Instrument \\ \hline
COSMOS  & \textit{u,g,r,i,z}                                                             & CFHT/MegaCam          \\
 &
  \textit{\begin{tabular}[c]{@{}l@{}}B, V, r$'$, i$'$, z$'$, \\ 12 medium-band optical\end{tabular}} &
  Subaru/Suprime-Cam \\
        & \textit{\begin{tabular}[c]{@{}l@{}}F606W, F814W, \\ F850LP\end{tabular}}       & \textit{HST}/ACS      \\
        & \textit{J1, J2, J3, H1, H2, K}                                                 & KPNO 4m/NEWFIRM       \\
        & \textit{J, H, Ks}                                                              & CFHT/WIRCam           \\
        & \textit{Y, J, H, Ks}                                                           & VISTA                 \\
        & \textit{\begin{tabular}[c]{@{}l@{}}F110W, F140W, \\ F125W, F160W\end{tabular}} & \textit{HST}/WFC3     \\
 &
  \textit{\begin{tabular}[c]{@{}l@{}}F090W, F105W, F115W, \\ F150W, F200W, F277W, \\ F356W, F410M, F435W, \\ F444W\end{tabular}} &
  \textit{JWST/NIRCam} \\
        & \textit{Channel 1,2,3,4}                                                       & \textit{Spitzer}/IRAC \\
        & \textit{F770W, F1800W}                                                         & \textit{JWST}/MIRI    \\ \hline
GOODS-S & \textit{U, R}                                                                  & VLT/VIMOS             \\
        & \textit{U38, B, V, R$_c$, I}                                                      & WFI 2.2m              \\
        & 14 medium bands optical                                                        & Subaru/Suprime-Cam    \\
 &
  \textit{\begin{tabular}[c]{@{}l@{}}F435W, F606W, F775W, \\ F814W, F850LP\end{tabular}} &
  \textit{HST}/ACS \\
        & \textit{\begin{tabular}[c]{@{}l@{}}F110W, F140W, \\ F125W, F160W\end{tabular}} & \textit{HST}/WFC3     \\
        & \textit{J, H, K$_s$}                                                              & VLT/ISAAC             \\
        & \textit{J, K$_s$}                                                                 & CFHT/WIRcam           \\
 &
  \textit{\begin{tabular}[c]{@{}l@{}}F090W, F105W, F115W, \\ F150W, F182M, F200W, \\ F210M, F277W, F335M,\\ F356W, F410M, F430M, \\ F435W, F444W, F460M, \\ F475W, F480M\end{tabular}} &
  \textit{JWST}/NIRCam \\
        & Channel 1, 2, 3, 4                                                             & \textit{Spitzer}/IRAC \\
        & \textit{F770W}                                                                 & \textit{JWST}/MIRI    \\ \hline
UDS     & \textit{U}                                                                     & CFHT/MegaCam          \\
        & \textit{B, V, R$_c$, i$_c$, z$'$}                                                      & Subaru/Suprime-Cam    \\
        & \textit{J, H, K$_s$}                                                              & UKIRT/WFCAM           \\
        & \textit{F606W, F775W, F814W}                                                   & \textit{HST/ACS}      \\
        & \textit{F125W, F140W, F160W}                                                   & \textit{HST/WFC3}     \\
 &
  \textit{\begin{tabular}[c]{@{}l@{}}F090W, F105W, F115W, \\ F150W, F200W, F277W, \\ F356W, F410M, F435W, \\ F444W\end{tabular}} &
  \textit{JWST/NIRCam} \\
        & Channel 1, 2, 3, 4                                                             & \textit{Spitzer/IRAC} \\
        & \textit{F770W, F1800W}                                                         & \textit{JWST/MIRI}   
\end{tabular}\label{tab:filters}
\end{table}

Photometry from DJA (v7.0 or v7.2) was measured within $0\farcs5$ apertures (\texttt{*\_tot\_1}), with total flux estimated using FLUX\_AUTO from \textsc{sep} \citep{Bertin1996, Barbary2016}.
Source detection was based on a weighted combination of all available NIRCam long-wavelength images (\textit{F277W, F356W} and \textit{F444W}), whereby the individual images were combined using their corresponding noise maps as weights \citep{Valentino2023}.
\hst\ photometry was corrected for flux outside the AUTO aperture. 
The remaining filters from the ground-based telescopes and \spitzer\ account for point-spread-function (PSF) variations and blending of nearby sources.

Given the heterogeneous angular resolution of the input imaging, we did not construct a fully PSF-matched catalogue.
Instead, we assessed the impact of PSF differences on the photometry by iteratively applying the \eazypy\ zero-point correction procedure to each band (five iterations). The resulting corrections are typically only a few percent, indicating that the adopted aperture size is sufficient to mitigate the effects of the varying PSFs when estimating photometric redshifts. Applying these corrections also improves the median absolute deviation (MAD) of the photometric redshifts by 10–30\%, depending on the field. We further find that including the available ground-based photometry reduces the MAD by up to a factor of two. The largest improvement is seen in the COSMOS field, where 39 additional photometric bands are available beyond the \jwst\ and \hst\ data, whereas the improvement in the other fields is typically <20\% owing to the smaller number of available ground-based bands.

For the photometric redshift fitting, we adopted the method and template set described in \cite{minju2024a} (\texttt{carnall\_sfhz\_13}).
The templates are corrected for Milky Way dust extinction based on the median RA and Dec of each field prior to fitting. Redshifts are allowed to vary between 0.01 and 15, with the upper limit informed by spectroscopic confirmations.
The right panel of Fig.~\ref{fig:redshift} compares photometric and spectroscopic redshifts, showing good agreement with a MAD of 0.017.
Deeper JWST photometry in the parent catalogue enables robust redshift estimates for fainter and more distant sources.

We note that our approach to estimating photometric redshifts and galaxy properties with \eazypy \,may not be optimal for every scientific application. We therefore encourage users to explore the released imaging data, including the \jwst\ and \hst\ photometry available through DJA, and to apply alternative analysis methods where appropriate, particularly for sources that may not be well represented by the adopted templates. Unless otherwise stated, the analyses presented in this paper are based on the \eazypy \,results described above.


\section{ALMA data retrieval, reduction, and imaging}\label{sec:dataprocessing}

We queried the ALMA archive using \textsc{astroquery} \citep{Ginsburg2019}, based on the parent catalogue's co-ordinates.
Figure \ref{fig:coverage} shows the current ALMA coverage (shaded grey regions) in the GOODS-S field, as an example, and the ALMA query area is shown as a dashed black line. 
Any ALMA project whose footprint intersects this query region is retrieved and included in our dataset.

In this first data release, and throughout the following description, we include only single-pointing (non-mosaic) ALMA observations.
We summarise the effective area in Table~\ref{tab:area} for each field, comparing with the DJA (catalogued) and 3D-HST.
The effective sky coverage of the reduced data as of
\rdate\, is $\approx$\surveyarea\ arcmin$^2$ where overlapping pointings are counted only once, and the footprint is defined by the region enclosed by the 20\% ALMA primary-beam response. For comparison, the summed sky coverage of all individual pointings, without removing overlaps, is $\approx$\totalarea\ arcmin$^2$.
The total number of reduced programmes is 220.
Within the ALMA footprint, \NgaltotALMA\ unique parent sources are covered. 
Of these, \NzspecALMA\, sources have spectroscopic redshifts (Table~\ref{tab:detnum}).

\begin{table}[]
\caption{Effective area covered by ALMA, JWST, and HST.}\label{tab:area}
\small
\begin{threeparttable}
\begin{tabularx}{\columnwidth}{X|ccc} \hline\hline
        & \multicolumn{3}{c}{Survey area (arcmin$^{2}$)} \\ \hline
Field   & ALMA (\ecogal)    & JWST (DJA)    & 3D-HST\tnote{a}    \\ \hline
COSMOS  & 151.6            & 267.5         & 183.9     \\
GOODS-S & 148.0            & 72.5          & 171.0     \\
UDS     & 91.4             & 265.4         & 191.2    
\end{tabularx}
\begin{tablenotes}
\footnotesize
\item[a] According to \citet{Skelton2014} 
\end{tablenotes}
\end{threeparttable}
\end{table}

The data reduction strategy follows that adopted by \acosmos\ \citep{Liu2019a} and \agds\ \citep{Adscheid2024}. 
We restored the calibrated MeasurementSet (MS) using the \texttt{scriptForPI.py} scripts by the Joint ALMA Observatory together with the raw archival data, within the Common Astronomy Software Applications package (CASA; \citealt{McMullin2007}).
The \texttt{scirptForPI.py} script reconstructs the calibrated MS by applying the calibration tables generated during the Quality Assurance Level 2 (QA2) process for each execution block. These include corrections for system temperature, antenna positions, water vapour radiometers, antenna gains, bandpass, amplitude, and polarisation. When required, additional calibration products, such as flux tables for calibrators, updated antenna positions, and supplementary flagging tables, are also applied for the calibration.

For datasets calibrated with the ALMA Science Pipeline, we used the CASA pipeline version specified in the corresponding QA2 report, or a compatible later version. The ALMA pipeline applies a set of heuristics to produce calibrated data products in a flexible and reproducible manner (see \citealt{casapipeline} for further details). In the vast majority of cases, the pipeline calibration produces satisfactory results, as confirmed by inspection of the calibrated images. Only when the calibrated images exhibit obvious calibration artefacts (e.g. striping) or other calibration-related issues do we reprocess the observations directly from the raw visibility data.
Two ALMA projects (2011.0.00064.S, 2022.1.00076.S) exhibited phase errors affecting a subset of calibrator scans. For these cases, we added a flag table and ran the reduction script prior to imaging.
For several early Cycle 0 datasets, we retrieved the already-calibrated measurement sets from ESO\footnote{\url{https://almascience.eso.org/local-news/requesting-calibrated-measurement-sets-in-europe}}, due to Python compatibility issues with the original calibration scripts.

\subsection{Imaging}\label{sec:imaging}
We generated continuum images using the CASA task \texttt{tclean}, adopting natural weighting and a \textsc{clean} threshold of $2\sigma$.
The \texttt{hogbom} deconvolver \citep{Hogbom1974} was used as the default \textsc{clean} algorithm.
Imaging parameters -- including stopping threshold, pixel scale, and image size -- were adjusted on a per-project and per-target basis according to the noise properties and dynamic range of the dirty maps. 
We specified the pixel scale and image size based on the dirty synthesised beam and the observing frequency. The pixel scale was chosen to sample the synthesised beam with at least five pixels, while the image size was set to cover twice the ALMA primary beam half-power beam width (HPBW). After applying the primary beam correction, we masked pixels with a primary beam response below 20\%, thereby defining the effective ALMA coverage area.

To determine the \textsc{clean} threshold in \texttt{tclean}, we estimated the noise level from the dirty image. We iteratively excluded circular regions around bright pixels above ($5\sigma$) and recalculated the noise until no significant ($>5\sigma$) peaks remain. The resulting noise estimate was then adopted as the final image noise level. We set the maximum number of iterations to ($\texttt{niter}=10000$), ensuring that the cleaning proceeds down to the target threshold of ($2\sigma$).
For a subset of early science programmes affected by backward-compatibility issues, imaging was performed with the CASA task \texttt{clean} (available in earlier CASA versions) with parameters manually configured to follow the $\texttt{tclean}$ strategy (e.g. natural weighting, a pixel scale sampling the synthesised beam with at least five pixels, and noise-based masking for \textsc{clean}).
We discuss the resulting impact of the homogeneous \textsc{clean} strategy described here in the next section.

We made collapsed 2D images without considering the line contamination using the multi-frequency synthesis \citep{Conway1990} imaging procedure without explicitly accounting for line contamination.
As discussed in Sect.~\ref{sec:linecontamination}, such contamination is negligible for the vast majority of detections.
In total, \nimage\ maps from single-pointing observations are considered in the first release, with a median spatial resolution of $\approx0\farcs7$.

\subsection{Photometry of ALMA data}\label{sec:photometry}
We measured the fluxes of ALMA-detected sources using three different approaches.
The first two relied on positional priors from the optical/NIR-detected parent catalogue, while the third was a blind extraction without any positional information.
Below, we describe the photometry procedure in detail.

\subsubsection{Photometry on prior positions}\label{sec:prior}

\begin{figure*}[h]
\centering
\includegraphics[width=0.98\textwidth]{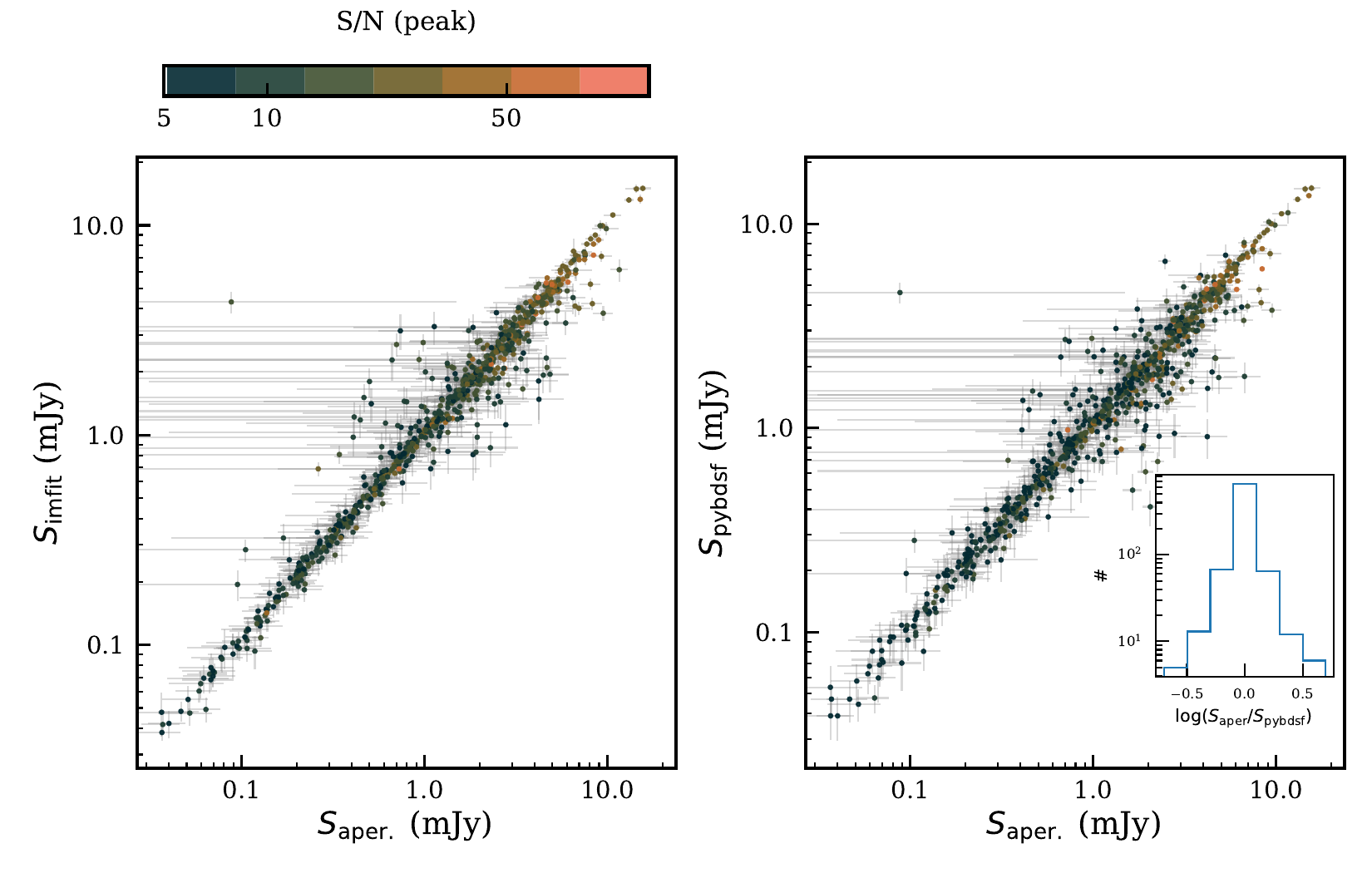}
\caption{Flux comparison for detected sources using the three extraction and measurement methods, colour-coded by S/N. Overall, the three methods show good mutual agreement.\label{fig:flux_all}}
\end{figure*}

For sources in the parent catalogue, we derived continuum flux densities using two approaches:
(i) fitting a two-dimensional (2D) Gaussian profile with the CASA task \texttt{imfit}, and
(ii) performing aperture photometry using apertures scaled to the synthesised beam.

We first searched for emission peaks with signal-to-noise ratios (S/Ns) above three within a radius of $r=0\farcs8$ around each prior position. To estimate the local noise level, we accounted for both the noise in the primary-beam-uncorrected image and the primary beam response at the source position using the following iterative procedure:
(i) Measure the noise (standard deviation) across the entire map.
(ii) Identify the brightest peak with $S/N>7$ and mask a region around it. The mask size is adaptive to the beam: for beams smaller than $0\farcs8$, we use a $0\farcs8$-diameter circular mask; for larger beams, we adopt a circular mask with radius equal to the beam major axis.
(iii) Recompute the noise excluding the masked regions.
(iv) Repeat the procedure until no $S/N>7$ peaks remain. The final noise is then scaled by the primary-beam response at the prior positions.

At the peak location above S/N>3, we fitted a 2D Gaussian model using \texttt{imfit}.
For practical reasons related to the \texttt{imfit} region-file requirement, we took a circular aperture size, which is adapted to the beam size; we take a $1\farcs0$-radius aperture size if the beam size is smaller than $1\farcs0$; otherwise, two times the major axis of the beam. 
We allowed all parameters of a single Gaussian component to vary freely during the fitting, including the amplitude, full width at half maximum (FWHM), and source position.
If the Gaussian fit converges successfully, we adopted the integrated flux density of the best-fitting Gaussian model.

We additionally performed aperture photometry using the same aperture size as adopted for the Gaussian fitting. If the peak S/N exceeds the initial detection threshold of peak S/N$>3$, we applied the aperture photometry. Otherwise, we report the 3$\sigma$ upper limit given the aperture sizes.

The uncertainties of both the \texttt{imfit} and aperture photometry measurements were estimated by accounting for the adopted aperture sizes; for \texttt{imfit}, we used the formulation from \citet{Condon1997}\footnote{\url{https://casadocs.readthedocs.io/en/latest/api/tt/casatasks.analysis.imfit.html}} taking into account the correlated noise within the apertures, whereas the aperture photometry error assumes uncorrelated noise. In Sect.~\ref{sec:sncut}, we define a data-informed detection threshold based on the pixel-based number statistics to produce a robust detection catalogue.

\subsubsection{Flux measurement validation and quality flags}

To validate the robustness of the extracted flux measurements, we compared the aperture and Gaussian-based flux estimates and examine the curve-of-growth profiles of the sources. 
The left panel of Fig.~\ref{fig:flux_all} shows the comparison with the prior-based aperture photometry ($S_{\rm aper}$) and the \texttt{imfit} results ($S_{\rm imfit}$) after cutting the peak S/N of \sncutprior\, (see Sect.~\ref{sec:sncut}).
The two measurements show overall good agreement; however, approximately $\approx$10\% of the sources display flux discrepancies larger than 30\%.

We examined the curve-of-growth profiles of the extracted sources. For a subset of sources, the emission is sufficiently extended and morphologically complex that a single Gaussian model does not fully recover the measured flux. Further investigation suggests that a subset of sources also exhibit decreasing curve-of-growth profiles, which are likely driven primarily by limitations in our uniform \textsc{clean} strategy. This effect is most prominent in the high-resolution datasets, where insufficient \textsc{clean} can leave residual negative sidelobes that reduce the cumulative flux measured within larger apertures. Re-imaging these fields with adaptive automasking and a deeper \textsc{clean} threshold of 1.5$\sigma$ substantially alleviates this behaviour for a subset of samples (Appendix~\ref{app:photometry}). A secondary contribution may arise from the reduced sensitivity of interferometric observations to extended emission in the highest-resolution data, although we expect this effect to be minor given the 2 arcsec aperture adopted for the photometry.

We therefore acknowledge that our uniform imaging strategy is not always optimal for all high-resolution datasets. In the first data release, we flag sources and add a \texttt{flag\_clean} column in the released catalogue whose flux measurements may be affected by \textsc{clean}-related imaging artefacts; these sources constitute approximately 8\% of the sample. Future data releases will revisit the imaging strategy for these datasets.
The vast majority of sources (>90\%) show consistent flux values from different photometry methods and different \textsc{clean} processes. We adopt aperture photometry as our baseline measurement for the scientific analysis presented in the following sections.

\subsubsection{Photometry on blindly detected sources}\label{sec:blind}
For a blind detection, we used the Python Blob Detector and Source Finder (hereafter \textsc{pybdsf}; \citealt{pybdsf}). 
We used images without primary-beam-attenuation correction as the detection map. 
\textsc{pybdsf} detection occurs based on the S/N, and the advantage of having the primary-beam-attenuation-uncorrected images is that the noise is held constant across the map.
The total continuum flux was corrected after detection is made with \textsc{pybdsf} based on the primary beam response using the \texttt{*.pb} file; the flux and the noise levels are boosted by 1/(pb response).
We activated the option of grouping by island (\texttt{group\_by\_isl=TRUE}) and use the following set-up, which was used in \citet{Liu2019a}; \texttt{thresh\_isl} = 3.0 and \texttt{thresh\_pix} = 4.0.
The island grouping option merges flux within a contiguous region that may contain multiple Gaussian components. 
Given the range of synthesised beam sizes, most sources are expected to be barely resolved, and this parameter is effective for resolved cases to get the total continuum flux.
The latter two parameters define the fitting regions (islands); sources with peak flux densities $\geq 4\sigma$ are searched, and the fitting region is demarcated by neighbouring pixels above the $3\sigma$ level.

The right panel of Fig.~\ref{fig:flux_all} shows the comparison between the aperture photometry and \textsc{pybdsf}-based measurements for those with optical/NIR counterparts and above peak $S/N>6.6$ (Sect.~\ref{sec:sncut}).
The distribution shows overall good agreement for most sources, although a slightly higher fraction of outliers is present ($\sim15\%$ with flux differences by 30\%) than \texttt{imfit}.
We conclude that all three detection methods and flux measurements perform reliably for the majority of the cases.

\begin{figure*}[ht]
\sidecaption
\includegraphics[width=12cm]{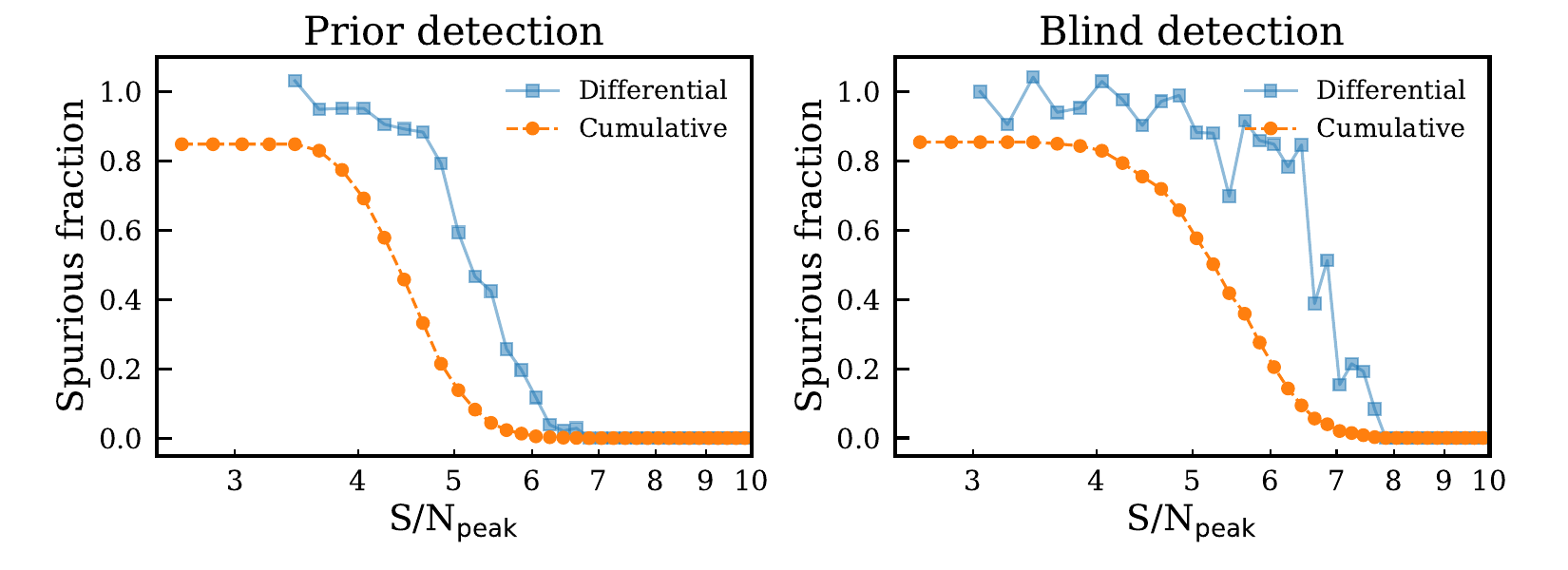}
\caption{Spurious fraction for the prior (left) and blind (right) source detections as a function of peak S/N. The cumulative spurious fraction indicates the fraction of spurious detections above a given $S/N_{\rm peak}$ threshold. We adopt detection thresholds of $S/N_{\rm peak}=\sncutprior$ for the prior-based catalogue and $\sncutblind$ for the blind catalogue, yielding a differential spurious fraction below $<$50\%.\label{fig:spurious}}
\end{figure*}

\subsection{Detection catalogue}\label{sec:detection}

\subsubsection{Astrometry between ALMA and optical/NIR}\label{sec:almaastrometry}
According to the ALMA technical handbook\footnote{\url{https://almascience.eso.org/documents-and-tools/cycle12/alma-technical-handbook}}, for observations using standard calibration and observing setups, the expected positional accuracy can be estimated as $\Delta = beam_{\rm FWHM}/S/N_{\rm peak}/0.9$, with $S/N_{\rm peak}$ the peak signal-to-noise ratio and $beam_{\rm FWHM}$ the synthesised beam.
For $S/N_{\rm peak}=3$, this corresponds to an accuracy of $0.37\times(beam)$.
We used this relation to evaluate the significance of the positional offsets.

For sources with $S/N>3$ and a plausible optical/NIR counterpart within one beam, the median expected ALMA positional accuracy is $0\farcs25$, given a median beam size of $1\farcs2$ and a median peak S/N of 3.9.
The measured offset to the optical/NIR counterpart is $0\farcs68$, exceeding the expected absolute positional accuracy of either dataset, though still within roughly half of the ALMA beam.

As expected, the positional offset depends on both S/N and beam size.
For sources with $S/N>10$, the median offset decreases to $0\farcs39$ with a median beam major axis of $0\farcs99$; the corresponding expected ALMA accuracy is $\Delta<0\farcs05$.
For high–angular-resolution ($<0\farcs.25$) data with peak $S/N>10 (30)$, the mean offset is $0\farcs11 (0\farcs12)$, compared to an expected positional accuracy of $\approx0\farcs01 (0\farcs004)$.
Thus, the measured offsets become increasingly significant relative to the expected absolute position accuracy at higher resolution and higher S/N.

These results suggest two points. 
First, global astrometric errors (between ALMA and optical/NIR) are negligible for source matching.
Second, the remaining offsets between ALMA and optical/NIR positions are likely driven by local effects rather than global systematics, though a detailed investigation is beyond the scope of this work.
Accordingly, in constructing the detection catalogue, we identify optical/NIR counterparts within a radius scaled to ALMA beam sizes for both the prior-based and blind-detection catalogues.

\subsubsection{S/N threshold based on inverted map}\label{sec:sncut}

Following the procedure of \citet{Liu2019a}, we determined an S/N threshold by comparing pixel-based statistics between the original map and its inverted counterpart.
The inverted map as obtained by multiplying the original map by -1, and sources were extracted from this map. 
The spurious fraction was then estimated as the ratio of detections in the inverted map to those in the original map, evaluated as a function of peak S/N.
Figure~\ref{fig:spurious} shows the resulting spurious fractions for both detection methods: prior-based detections using the parent catalogue with the aperture photometry and blind detections using \textsc{pybdsf}.

Based on this, we applied an $S/N_{\rm peak}$ cut at \sncutprior$\sigma$ and \sncutblind$\sigma$ to our prior-based and blind source detection catalogues, respectively. 
These thresholds are chosen to have the differential spurious fraction become $\lesssim50\%$ at the applied S/N$_{\rm peak}$.
The cumulative spurious fractions are $<6\%$ and $<10\%$ for the blind and prior-selected samples, giving similar spurious fractions compared to the \acosmos\, in \citet{Liu2019a}.

Since sources are not always compact, a more complete assessment would ideally involve field-dependent S/N thresholds derived from injecting and recovering artificial sources spanning a range of flux densities and angular sizes. However, a detailed analysis by \citet{Liu2019a} showed that completeness strongly depends on source size. In particular, defining detection thresholds based on the total-flux S/N can preferentially select compact sources, whereas an S/N
peak-based criterion provides a more robust balance between completeness and the spurious detection rate. Performing dedicated source-injection and recovery simulations for each field would therefore require a more extensive analysis of the completeness properties of the heterogeneous dataset, which is beyond the scope of the systematic approach adopted in this work.

\subsubsection{Positional offset threshold and detection catalogue}\label{sec:detectioncatalogue}

We considered source misidentification, especially for larger beams, first by taking into account the assessment of the astrometry and the offset described in Sect.~\ref{sec:almaastrometry}. 
The median offset is $\approx$60\% of the beam size; hence, we masked out sources with offset $>0.6\times$ the beam size if the beam size is greater than $0\farcs25$.
For high-resolution observations, we loosened the criteria to accommodate the possible small-scale positional offset between the dust and optical/NIR emissions within galaxies, as indicated by high S/N ($>10$), high-resolution ($0\farcs25$) observations.
We used the offset threshold of $<0\farcs2$ for high-resolution observations because sources with high peak S/N ($>30$) with beam sizes $<0\farcs25$ can have offsets up to $0\farcs20$.

Applying the peak S/N threshold and positional-offset cut to the prior-based catalogue yields \NalldetpriorRes\ detections. Some of these detections correspond to the same astrophysical sources observed in multiple programmes, through deeper integrations, or in multiple ALMA bands. After removing these overlapping observations, we obtain a sample of \NalldetpriorResUniq\ unique sources, of which \NalldetpriorResUniqz\ have spectroscopic redshifts.
For the blind catalogue, \Nalldetblind\, detections qualify the blind S/N threshold. 
Of these, \Nalldetblindopt\ detections (corresponding to \Nuniqdetblindopt\ unique sources) have an optical/NIR counterpart that meets the same offset requirement used for the prior catalogue.
To construct the secure source catalogue, we combined the prior and blind detections after cross-matching to optical/NIR counterparts. 
We summarise the number of sources that are in the parent catalogue, with the ALMA coverage, and detected in Table~\ref{tab:detnum} and in Table~\ref{tab:bands} for the breakdown of the number of sources for each ALMA band.

\begin{table}[]
\caption{Number of sources in the parent and the \ecogal\ catalogues.\label{tab:detnum}}
\small
\begin{threeparttable}
\begin{tabularx}{\columnwidth}{Xcc}
\hline \hline
        & Number       & 
         \begin{tabular}[c]{@{}l@{}}Number \\ with spec-$z$\end{tabular}\\ \hline
\begin{tabular}[c]{@{}l@{}}Parent sample in CANDELS \\ (COSMO, UDS, GOODS-S)\end{tabular} & \Ntotparent                                                     & 
\begin{tabular}[c]{@{}l@{}}\NzspecSecure\\ (\NzspecALL)\tnote{e}\end{tabular}
  \\ \hline \hline

\multicolumn{3}{c}{Data release 1 (version \rdate)} \\ \hline

\begin{tabular}[c]{@{}l@{}}Parent sample in the ALMA footprint\\\end{tabular}    & \NgaltotALMA                                                     &        \NzspecALMA         \\ \hline
\begin{tabular}[c]{@{}l@{}}Prior-based detection\\ ($S/N_{\rm peak}$\textgreater{}\sncutprior)\end{tabular} &
  \begin{tabular}[c]{@{}l@{}}\NalldetpriorRes\tnote{a}\\ (\NalldetpriorResUniq\tnote{b} )\end{tabular} &
  \NalldetpriorResUniqz\tnote{b} \\ \hline
\begin{tabular}[c]{@{}l@{}}Blind detection\\  ($S/N_{\rm peak}$\textgreater{}\sncutblind)\end{tabular}    & \Nalldetblind\tnote{a}                                                        &      226\tnote{b,c}    \\ \hline
\begin{tabular}[c]{@{}l@{}}Blind detection \\ without opt/NIR counterpart\end{tabular}    & 
\begin{tabular}[c]{@{}l@{}} \Nblindnocounterpart\tnote{a}   \\ (\Nblindnocounterpartuniq\tnote{b,d} )\end{tabular}                                                       &                 \\ \hline
\begin{tabular}[c]{@{}l@{}}Prior or Blind\\ (union of samples)\end{tabular}
& \begin{tabular}[c]{@{}l@{}}\Nall\tnote{a}\\ (\Nalluniq\tnote{b} )\end{tabular} & \begin{tabular}[c]{@{}l@{}} \Nallz\\ (\Nalluniqz\tnote{b} )\end{tabular}   \\ \hline
\begin{tabular}[c]{@{}l@{}}Prior and Blind\\ (intersection of samples)\end{tabular} &
\begin{tabular}[c]{@{}l@{}}\NSecure\tnote{a}\\ (\NuniqSecure\tnote{b} )\end{tabular}  & 224\tnote{b}   
\end{tabularx}
\begin{tablenotes}
\footnotesize
\item[a] Including overlap source positions from different bands.
\item[b] Unique sources are counted after merging duplicate detections across bands.
\item[c] Among 408 (unique) optical counterparts.
\item[d] Position tolerance of $0\farcs2$.
\item[e] Including 3D-HST grism spectroscopy
\end{tablenotes}
\end{threeparttable}
\end{table}

\begin{table*}[t]
\caption{Number of images and detections per ALMA band.\label{tab:bands}}
\centering
\small
\begin{tabularx}{\textwidth}{l*{7}{>{\centering\arraybackslash}X}}
\hline \hline
            & Band 3 & Band 4 & Band 5 & Band 6 & Band 7 & Band 8 & Band 9 \\ 
\midrule
Number of images & 268    & 227    & 9      & 400    & 996    & 16     & 17     \\
\begin{tabular}[c]{@{}l@{}}Number of prior galaxies within \textgreater 20\% \\ALMA primary response\end{tabular} &
  \begin{tabular}[c]{@{}l@{}}125,259\\ (87,661)\end{tabular} &
  \begin{tabular}[c]{@{}l@{}}62,942\\ (46,048)\end{tabular} &
  \begin{tabular}[c]{@{}l@{}}2626\\ (2277)\end{tabular} &
  \begin{tabular}[c]{@{}l@{}}30,144\\ (25,903)\end{tabular} &
  \begin{tabular}[c]{@{}l@{}}36,529\\ (29,114)\end{tabular} &
  \begin{tabular}[c]{@{}l@{}}735\\ (712)\end{tabular} &
  \begin{tabular}[c]{@{}l@{}}220\\ (210)\end{tabular} \\
\begin{tabular}[c]{@{}l@{}}Prior-based detection ($S/N_{\rm peak}$\textgreater{}5.1)\end{tabular} & 111 & 242 & 6 & 309 & 565 & 14 & 15 \\
\begin{tabular}[c]{@{}l@{}}Blind detection ($S/N_{\rm peak}$\textgreater{}6.6)\end{tabular}       & 62  & 133 & 6 & 245 & 598 & 13 & 13 \\
\bottomrule
\end{tabularx}
\end{table*}

\begin{figure}
\centering
\includegraphics[width=0.48\textwidth]{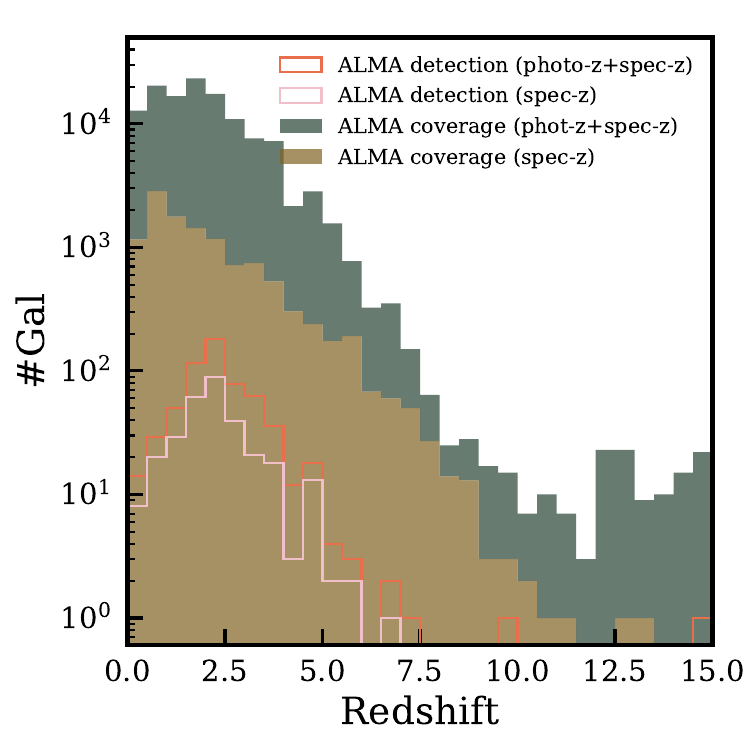}
\caption{Redshift distribution of ALMA-covered sources from the parent catalogue and of ALMA-detected sources. ALMA-detected galaxies are mainly located at $z\sim2.5$. The two ALMA detections with photometric redshifts $z>9$ are likely lower-redshift sources missing photometry at $>2\,\mu$m in the parent catalogue.\label{fig:zdist_det}}
\end{figure}

The merged detection catalogue contains \Nall, representing \Nalluniq\ sources that satisfy at least one of the two detection criteria.
Of these detections, \Nallz\ have spectroscopic redshifts, corresponding to \Nalluniqz\ unique sources in the detection catalogue.
Figure~\ref{fig:zdist_det} presents the redshift distribution of the ALMA-detected galaxies compared to the parent sample.
The majority of the sources ($239/\Nalluniqz \approx77\%$) lie at $1<z<3$, with additional detections extending to higher redshift: 39 galaxies at $3<z<4$, and 21 at $z>4$. Table~\ref{tab:redshift} presents the redshift breakdown of the parent sample with ALMA coverage and ALMA-detected sources for each field.

\begin{table}[]
\caption{Redshift distribution of parent galaxies and ALMA detections.\label{tab:redshift}}
\centering
\small
\begin{tabularx}{\columnwidth}{l*{6}{>{\centering\arraybackslash}X}}
\hline\hline
 & 0\textless{}z\textless{}0.5 & 0.5\textless{}z\textless{}1 & 1\textless{}z\textless{}2 & 2\textless{}z\textless{}3 & 3\textless{}z\textless{}4 & z\textgreater{}4 \\ \hline
\multicolumn{7}{l}{Parent galaxies with ALMA coverage} \\ \hline
ALL      & 1161  & 2850  & 3207  & 1890  & 1276 & 1153 \\
COSMOS   & 601   & 1158  & 1091  & 686   & 378  & 396  \\
GOODS-S  & 426   & 1206  & 1475  & 834   & 642  & 541  \\
UDS      & 134   & 486   & 641   & 370   & 256  & 216  \\ \hline
\multicolumn{7}{l}{ALMA detection}                     \\ \hline
ALL      & 8     & 20    & 91    & 128   & 39   & 21   \\
COSMOS   & 4     & 8     & 26    & 43    & 11   & 11   \\
GOODS-S  & 4     & 8     & 43    & 39    & 15   & 2    \\
UDS      & 0     & 4     & 22    & 46    & 13   & 8   
\end{tabularx}
\end{table}

To assess whether ALMA detections represent a random subset of ALMA-covered galaxies, we performed a Kolmogorov–Smirnov (KS) test.
The result ($D=0.32$ with $p=2\times10^{-27}$) rejects the null hypothesis, indicating that the detected sample is not drawn randomly from the full ALMA-covered population.
This reflects both the heterogeneous depths of the observations and a genuine selection effect: ALMA observations are made biased towards dusty galaxies, whose redshift distribution peaks near the epoch of maximum cosmic star formation activity.

A total of \Nblindnocounterpart\ (corresponding to \Nblindnocounterpartuniq\ unique positions, using an offset tolerance of $0\farcs2$) sources do not have an optical counterpart in the parent catalogue.
Of these, 44\% have a peak S/N $<8$ and may be spurious at $\approx50\%$ chance. 
Some lie outside the \jwst\ coverage used to construct the DJA catalogue (v7.0, v7.2), where the F160W filter from \hst\ is the longest wavelength available.

As a crude check on whether the number of detections is reasonable, we treat these sources as potential H-dropout (i.e. F160W-dropout) galaxies.
Assuming that they lie at $3<z<5$ (e.g. \citealt{Wang2019c, Talia2021}) and accounting for the survey area and a $\sim50\%$ spurious fraction, we obtain a number density of  $n\sim2\times10^{-5}$ Mpc$^{-3}$ (without completeness corrections). 
This is broadly consistent with the expected number density of H-dropout galaxies, $n\sim1-2\times10^{-5}$ Mpc$^{-3}$ for $3.5<z<5.5$ (e.g. \citealt{Wang2019c}, see also \citealt{Talia2021} and references therein, for UV-dark galaxy selections).

However, we note that the true number may vary depending on survey completeness and sample selection. 
For \ecogal, which has heterogeneous depth and coverage, completeness corrections are challenging.
Additionally, the submillimeter selection may not fully overlap with H-dropout selections (e.g. \citealt{Long2026}).
A detailed investigation is beyond the scope of this paper; however, several sources lack a clear counterpart even in F444W, warranting further study.
The nature of one such source is discussed in Sect.~\ref{sec:optdark}.

\subsubsection{Line contamination assessment}\label{sec:linecontamination}
Continuum images were created by collapsing all channels of all spectral windows (SPWs) (Sect.~\ref{sec:imaging}).
However, the original programmes may target the line emissions.
Strong emission lines such as \cii\, and CO lines can impact the underlying continuum flux measurements.
The median bandwidth of the retrieved ALMA programme is the nominal $\sim8$ GHz, as Fig.~\ref{fig:bandwidth} shows. 
Depending on redshift and the line strength, which typically correlates with the SFR, the line contribution can become significant and, in extreme cases, dominate the measured flux.
We therefore estimate the potential line contamination as described below.

\begin{figure}
\centering
\includegraphics[width=0.48\textwidth]{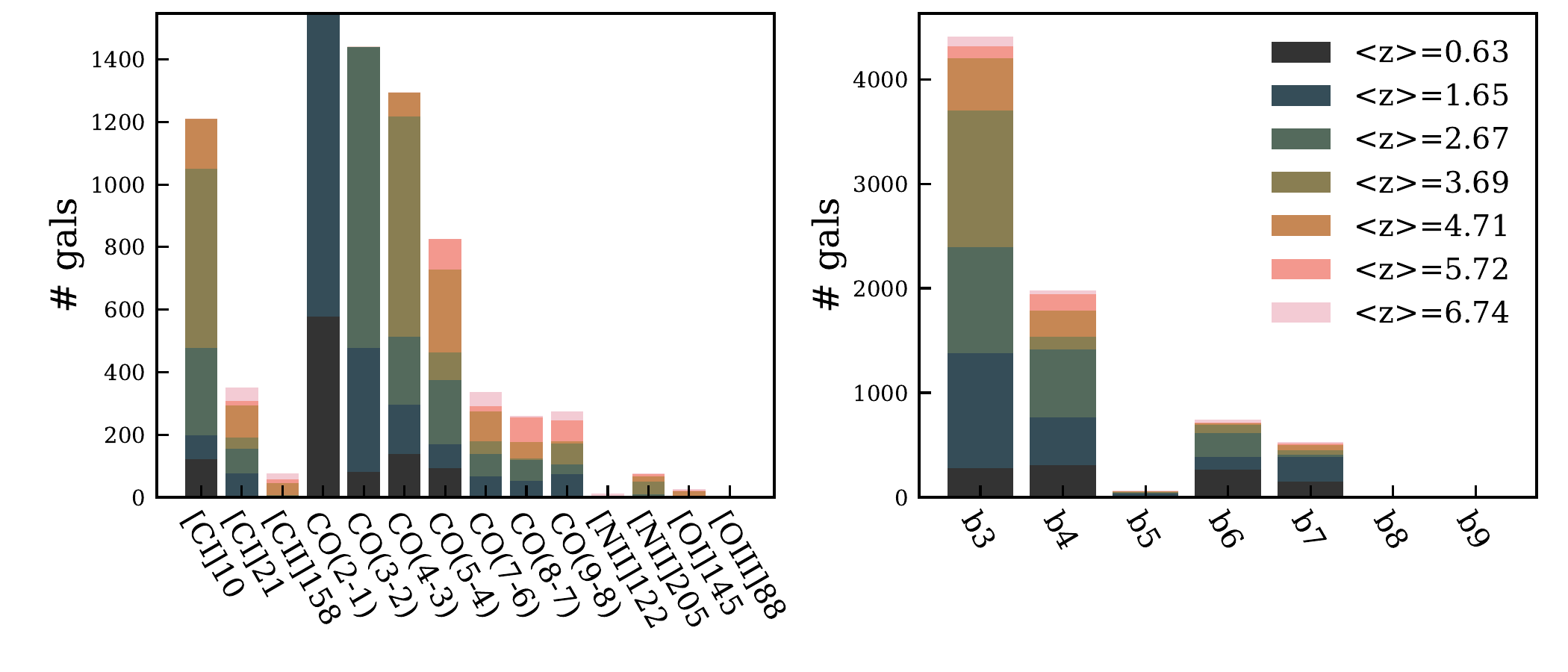}
\caption{Stacked histogram of the spectral lines that fall within the ALMA bandwidth for each redshift bin, shown for the spectroscopically confirmed sample.\label{fig:targetedline}}
\end{figure}

As a first step, we examined which spectral lines would fall within the accessible ALMA bands at different redshifts, taking into account both the redshift and the bandwidth coverage.
Figure~\ref{fig:targetedline} presents stacked histograms of the targeted lines (left) and the corresponding ALMA bands (right) for the spectroscopically confirmed sample.
This histogram shows: (i) At low redshift ($z<1$), the ALMA observations predominantly target low-$J$ CO transitions
(ii) At higher redshift ($z>4$), the targets shift towards higher-$J$ CO lines ($J > 5$) and the \cii\ 158 $\mu$m line, as these transitions move into ALMA’s most favourable atmospheric windows.
The band distribution (right panel of Fig.~\ref{fig:targetedline}) indicates that Bands 3 and 4 cover more galaxies, reflecting their wider field of view and velocity coverage.

\begin{figure}
\centering
\includegraphics[width=0.48\textwidth]{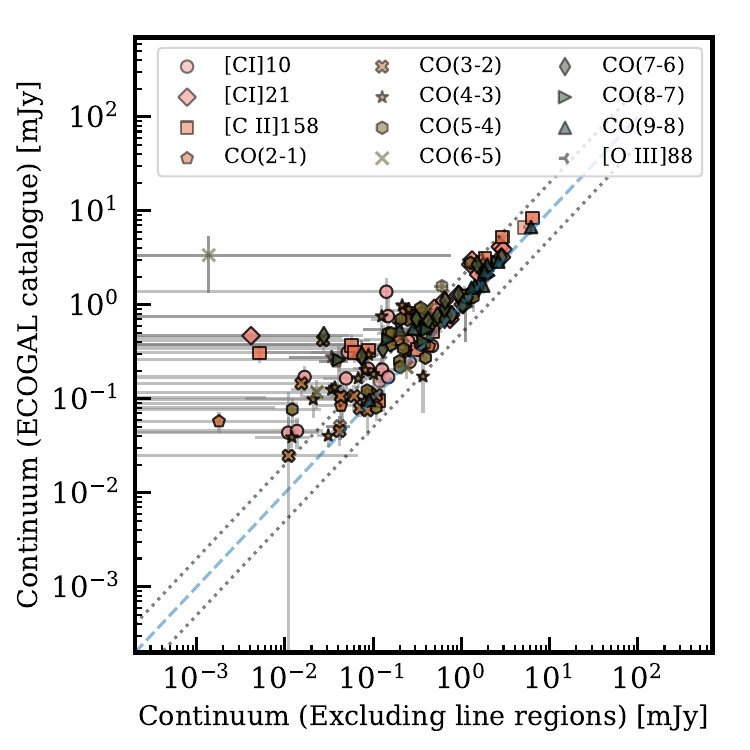}
\caption{
Comparison of full-SPW continuum fluxes and continuum fluxes remeasured with line-emitting channels excluded, for spectroscopically confirmed, dust-detected, line-detection sources. Overall agreement indicates that line contamination is not a dominant source of the flux from the fully collapsed image. The dashed lines mark flux ratios of 0.5 and 2 between the original and line-excluded measurements. \label{fig:linefree}}
\end{figure}

We further restricted our analysis to continuum-detected sources that satisfied the S/N threshold (Sect.~\ref{sec:sncut}) and then assessed which of the spectroscopically confirmed galaxies could have potential line coverage. 
Among the \Nallz\ detections (from either the blind or prior catalogues; see Table.~\ref{tab:detnum}) with optical counterparts and redshift -- counting duplicates when the same source is observed in multiple bands -- we identify 202 cases ($\sim$30\%) in which lines might have been targeted. 
The remaining observations target the continuum only.
For these line-covered cases, we re-measured the continuum fluxes after masking the expected line region based on the spectroscopic redshift.

Figure~\ref{fig:linefree} shows the comparison of the continuum fluxes with and without line regions excluded.
Among them, we find 96 cases in which line emission contributes more than 20\% in their continuum maps, corresponding to $\approx$14\% of spectroscopically confirmed samples. 
The fraction is roughly $\approx2\times$ higher than the estimate from \citet{Liu2019a}, where they estimated 7\% of their sample would have a contribution of more than 20\% in the continuum; a higher fraction is expected here because our analysis is restricted to spec-$z$ confirmed sources.
Among these 96 cases, 63 are strongly line-dominated in the continuum, with line contributions exceeding 50\% of the measured continuum.
Figure~\ref{fig:strongline} presents two extreme cases where strong lines dominate the continuum measurement within the ALMA bandwidth.

However, we expect that the dominance of strong lines over the continuum is not the majority of cases for continuum detection.
When the observation is not designed to target the line emission, there is a low chance (e.g. $\sim1.6\%$ by \citealt{Cooke2018}) of a line that could contaminate the continuum flux.
Provided that the typical emission line width of galaxies is 300 km s$^{-1}$, the line would occupy a few percent of the entire bandwidth, while higher occupancy is expected in higher frequency bands and/or high redshift sources.
Therefore, we opt to use the collapsed measurements for the first catalogue and data release. 
In the first data release, we add a flag column for the line contamination flag (\texttt{flag\_line}) and the reestimated continuum flux (\texttt{flux\_linefree\_aper}), excluding line-emitting regions for spectroscopically confirmed, line-detection cases discussed above.


\section{Galaxy properties overview}\label{sec:galaxyproperties}
\subsection{UVJ colour-colour space and star-forming main sequence}

To gain an insight into the types of galaxies detected in the ALMA bands, we used the stellar mass and the SFR of the source using the output from \eazypy.
It would be ideal, and we encourage readers to explore different spectral energy distribution (SED) fitting tools, depending on their specific scientific purposes.
However, for a general context and simplicity, we stick to using the output from \eazypy\footnote{The stellar mass is the most robust parameter for galaxies $z<6$ with rest-frame optical data points available from \jwst.
As shown in Fig.~\ref{fig:zdist_det}, the majority of sources, especially ALMA-detected sources, are located below $z<6$. 
We ran \textsc{fast++} (C++ version of \textsc{FAST}, \citealt{Kriek2009, Schreiber2018b}, \url{https://github.com/cschreib/fastpp}) for the spec-$z$ confirmed galaxies and confirmed that the stellar mass from \eazypy\ is reasonably in agreement within 0.28 dex (median value), where \eazypy\ stellar mass is higher. This has been known in previous studies (e.g. \citealt{Gould2023, Ito2026}).}.
In the next section, we discuss the uncertainty associated with the SFR from \eazypy\ and how \ecogal\, can address the issue.

\begin{figure*}
\centering
\includegraphics[width=0.98\textwidth]{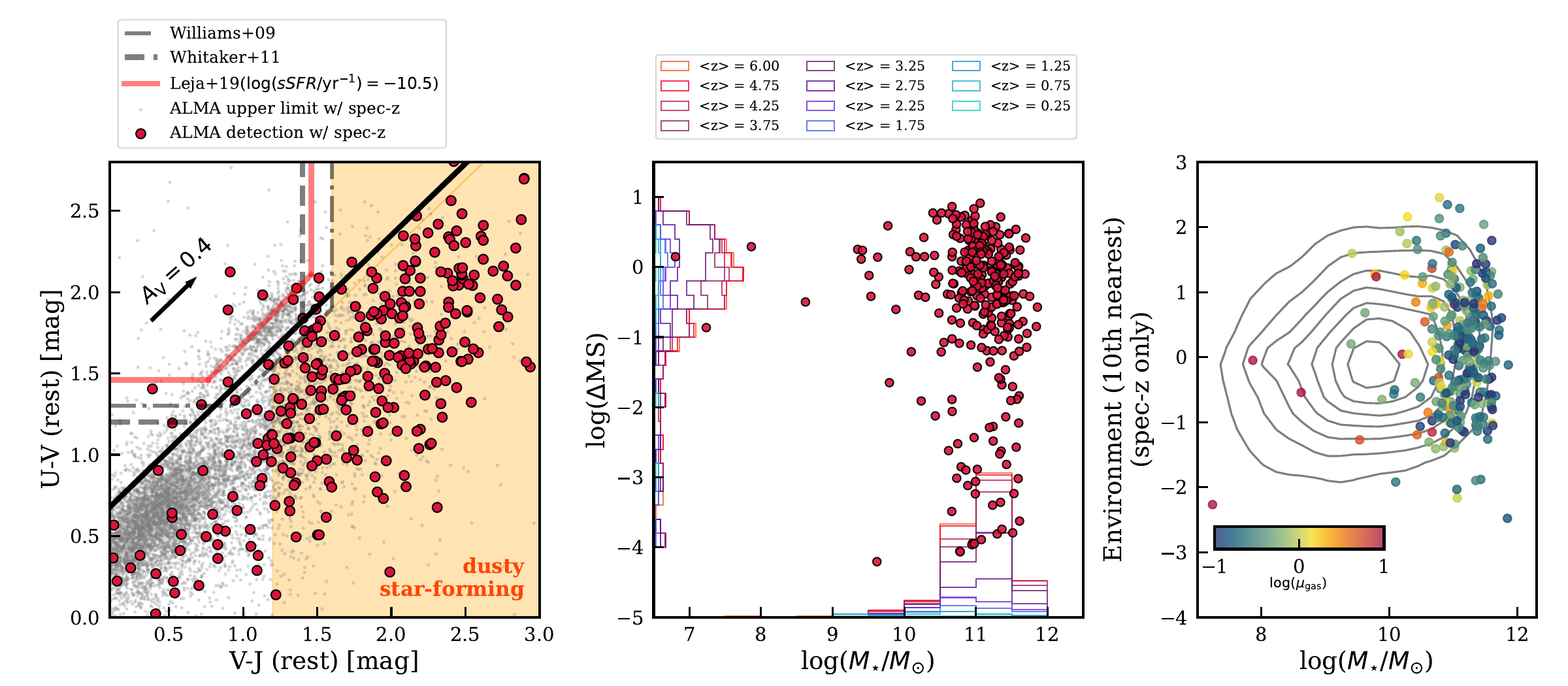}
\caption{Parameter space of \ecogal\ galaxies with spectroscopic redshift. Galaxy parameters (UVJ colours, SFR, and stellar mass) are derived from \eazypy. Left: UVJ colour–colour diagram. ALMA-detected sources are shown as larger circles, colour-coded by redshift, while small grey dots represent galaxies with ALMA upper limits. Most sources lie in the star-forming region, with many having $V-J >1.2$, corresponding to $A_{\rm V}\sim1.6$ for constant SFR models and age $<$1 Gyr; e.g. \citealt{Spitler2014}). Middle: Offset from the star-forming main sequence ($\log{(\Delta{\rm MS})}$) versus stellar mass for ALMA-detected sources, with the histogram showing the redshift distribution. Right: 3D local galaxy density (based on the tenth-nearest neighbour) as a function of stellar mass, with colour indicating gas fraction ($\log(\mu_{\rm gas})  = \log{(M_{\rm gas}/M_{\star}})$) (Sect.~\ref{sec:mugas}).\label{fig:galparams}}
\end{figure*}

The left panel in Fig.~\ref{fig:galparams} shows the distribution of \ecogal\, galaxies in the UVJ colour diagram based on the prior catalogue.
As expected, most of ALMA-detected sources occupy the star-forming region of the UVJ diagram. 
This naturally reflects ALMA’s sensitivity to far-infrared (FIR) dust emission, which is most prominent in actively star-forming galaxies.
Only a small fraction of ALMA-detected galaxies (2-4\%, depending on the different colour definition, e.g. \citealt{Williams2009, Whitaker2011, Leja2019b}) fall in the UVJ-quiescent region.

The middle panel of Fig.~\ref{fig:galparams} shows the deviation from the main sequence ($\log(\Delta$MS)), using the \citet{Speagle2014} relation, as a function of stellar mass for ALMA-detected, spectroscopically redshift-confirmed galaxies in the prior catalogue.
Most sources lie within 1 dex of the main sequence and are therefore star-forming, typically at the massive end ($\gtrsim10^{10}\,M_{\odot}$).
The stacked redshift histogram in the same panel further shows that the bulk of the detections lie at $z\sim2$, as shown in Fig.~\ref{fig:zdist_det}.

We find that 6\% (27/307) of galaxies in our spectroscopic sample lie significantly above the star-forming main sequence ($>$0.6 dex; hereafter referred to as starburst galaxies). The starburst fraction peaks near cosmic noon, reaching approximately 20\% at $1.5<z<2.0$, while remaining below 10\% in all other redshift bins. At $z>4$, the spectroscopic sample is limited to 21 galaxies, of which only one satisfies the adopted starburst criterion.

The evolution of the starburst fraction has been extensively investigated since the launch of Herschel. Restricting comparisons to massive galaxies ($M_{\star} > 10^{10}$ M$_{\odot}$), \citet{Rodighiero2011} reported that only 2-3\% of galaxies at $z\approx2$ qualify as starbursts using a comparable definition, namely galaxies with SFRs elevated by more than a factor of four (0.6 dex) above the star-forming main sequence. 
Similarly, 
 \citet{Schreiber2015} found a nearly constant starburst fraction of $2-4\%$ over the redshift range $0 < z < 4$. 
In contrast, \citet{Bisigello2018} measured higher fractions ($\approx12-16\%$) at $1 < z < 3$, likely reflecting differences in the adopted starburst definition, which was based on an sSFR threshold and included galaxies with lower stellar masses.
At higher redshift, 
\citet{Caputi2017} reported that fewer than 5\% of massive galaxies ($M_{\star} \gtrsim 2 \times 10^{10}$ M$_{\odot}$) at $z \approx 4-5$ are classified as starbursts using an sSFR criterion, although a substantially larger starburst fraction was found among lower-mass galaxies (see also \citealt{Rinaldi2025}).

In this context, the relatively high starburst fraction measured at $1.5 < z < 2.0$ is likely influenced by the selection function of ALMA archival observations, which preferentially target galaxies with elevated star formation activity; the bias is also supported by our assessment of the redshift distribution of the ALMA continuum-detected sources discussed in Sect.~\ref{sec:detectioncatalogue}. 
An unbiased measurement of the starburst fraction at these redshifts will require different analysis methods, such as removing the central source contributions (e.g. \citealt{Bollo2025}), or larger, more representative samples from wide-area surveys, with extensive spectroscopic follow-up.

Meanwhile, the fraction of galaxies appearing quiescent based on their sSFR (12\% with $\Delta{\rm MS}<-2$ dex) is higher than the fraction classified as quiescent via UVJ colours.
Earlier studies have shown that different selection methods typically yield different quiescent galaxy selections (e.g. see the Venn diagram in \citealt{Ito2026}).
In the context of ALMA-selected samples, it likely reflects heavily dust-obscured galaxies whose SFRs are underestimated from \eazypy\ (Sect.~\ref{sec:sfr}).

\subsection{Star formation rate from dust, H$\alpha$, and SED}\label{sec:sfr}
As part of \ecogal's extended data products, which incorporate spectroscopic information from the DJA, we provide measurements enabling detailed studies of star formation activity using multiple tracers. 
Here we demonstrate these capabilities using the subset of galaxies that have both an ALMA detection (from the prior-detection catalogue) and H$\alpha$ and H$\beta$ measurements from \jwst/NIRSpec.

\jwst\,'s wavelength coverage enables studies of redshifted H$\alpha$ line emissions for $z>3$ galaxies, opening new possibilities for studying star formation from rest-frame optical lines.
Further, when the H$\beta$ line is simultaneously obtained, dust attenuation corrections can be made via the Balmer decrement (e.g. \citealt{Berman1936, Calzetti1994, Groves2012, Reddy2015, Shapley2023, Matharu2024, Sandles2024}).

Among the \NzspecALMA\ spectroscopically confirmed galaxies (Table~\ref{tab:detnum}) that have ALMA coverage, \NzDJAALMA\ have JWST/NIRSpec spectra.
The emission line flux is extracted using the \texttt{msaexp} \citep{msaexp} reductions provided in the DJA. 
Details of the line analysis are described in \citet{Ito2026}.
In what follows, we focus on the NIRSpec spectra of galaxies in the prior ALMA detection catalogue.

\begin{figure}
\centering
\includegraphics[width=0.48\textwidth]{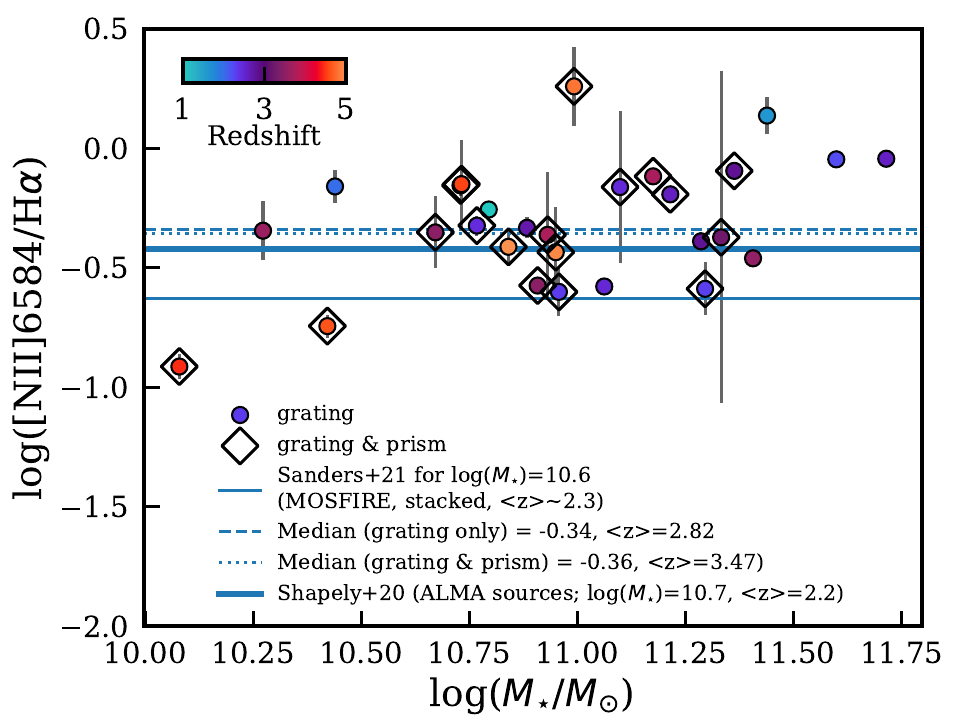}
\caption{Line ratio between \nii$\lambda6584$ and H$\alpha$ for 28 ALMA-detected galaxies, colour-coded by redshift. The median ratios for subsets of the sample (grating-only and PRISM+grating) are higher than those reported by \citet{Sanders2021} for $z\approx2.3$ galaxies but comparable with \citet{Shapley2020}, which studied ALMA sources at $z\approx2.2$. (Sect.\ref{sec:niicorr}).\label{fig:n2ratio}}
\end{figure}

\begin{figure*}[thb]
\centering
\includegraphics[width=0.95\textwidth]{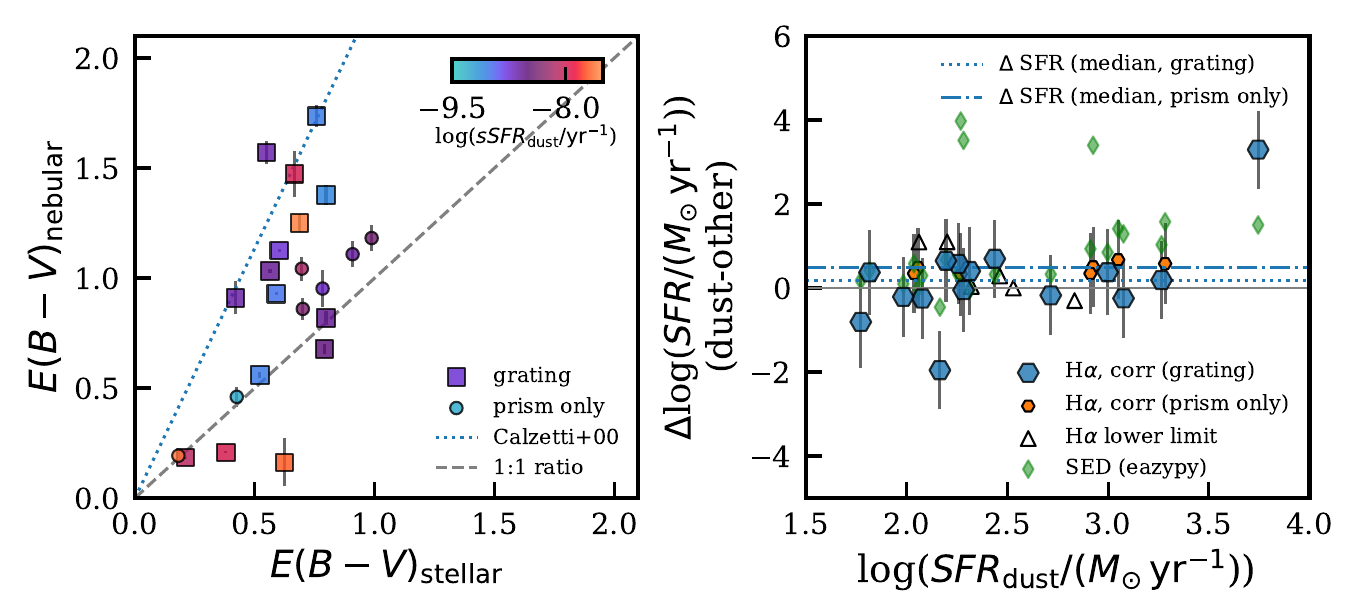}
\caption{Left: Nebular colour excess ($E(B-V)_{\rm neb}$) dervied from H$\alpha$ and H$\beta$ fluxes using Eq.~\ref{eq:ebv}, colour-coded by dust-based specific star formation rate (sSFR). For reference, the \citet{Calzetti2000} relation ($E(B-V)_{\rm neb} = 2.27\times E(B-V)_{\rm stellar}$), and a 1:1 ratio are shown. Right: Comparison of $\log({\rm SFR})$ between dust-based SFR versus H$\alpha$ (Balmer-decrement corrected, hexagons) and SED fitting (from \eazypy). Larger hexagons indicate \jwst/NIRSpec grating observations, while smaller hexagons correspond to PRISM-only measurements. \label{fig:sfr}}
\end{figure*}

We obtained the dust-corrected SFR from H$\alpha$ with the following procedure. 
We first extracted the emission-line fluxes from the DJA catalogue (v4.3, e.g. \texttt{line\_ha, line\_ha\_nii, line\_nii\_6549, line\_nii\_6584 line\_hb}) and the associated errors. 
When multiple measurements are available for the same \texttt{objid}, we select the spectrum with the higher S/N. 
This occurs because multiple programmes may target the same object, sometimes independently. This is particularly common for MSA observations, where filler targets can be duplicated across different masks, resulting in spectra with different S/N due to varying exposure depths. Although combining these spectra would in principle improve the S/N, the current DJA processing treats individual observations independently rather than producing a science-specific combined catalogue. For consistency, we also do not combine ALMA observations of the same source obtained at different depths in this first data release. Instead, we leave the combination of individual observations to users depending on their specific scientific goals.

We required S/N $>$1 for all lines (H$\alpha$, H$\beta$ and \nii$\lambda\lambda$6549, 6584 when available), yielding a sample of 48 galaxies.
Of these, (i) 38 have PRISM-based H$\alpha$ (blended with \nii) and H$\beta$ (PRISM or grating) (hereafter prism sample), (ii) 28 sources have grating-based H$\alpha$ and H$\beta$ (PRISM or grating) (hereafter grating sample).
18 sources overlap between these two groups.
The redshifts span $z=[1.60, 6.85]$ (prism) and $z=[1.04, 5.22]$ (grating) with the median values $z=3.16$ and $z=2.75$, respectively. 

The adopted S/N threshold ensures that all selected sources have S/N$>3$ in either H$\alpha$ (for the grating sample) or the combined H$\alpha$+\nii$\lambda\lambda$6548,6583 flux (for the prism sample). 
This provides a robust anchor for the line-ratio measurements used in the subsequent analysis. For the grating sample, 23/28 sources also have S/N$>3$ detections of \nii$\lambda\lambda$6548,6583, demonstrating that the \nii/H$\alpha$ ratios used to correct the prism fluxes (Sect.~\ref{sec:niicorr}) are generally robust against the adopted selection threshold. 
As expected, H$\beta$ is the limiting line due to its lower intrinsic flux, with 20 sources having S/N$>3$ detections in all four lines and 28 sources retaining lower-significance H$\beta$ measurements with $1<$S/N$<3$.

\subsubsection{Correction for \nii\, contribution in PRISM and slit loss}\label{sec:niicorr}

NIRSpec/PRISM resolution ($R < 300$ at shorter wavelengths) blends H$\alpha$ with the \nii\ doublet (\nii$\lambda\lambda6549, 6584$) for $z\lesssim6$.
Therefore, we need to correct the PRISM H$\alpha$ measurements using constraints from the grating spectra.

The \nii/H$\alpha$ ratio primarily traces metallicity and ionisation conditions (e.g. \citealt{Kewley2002, Tremonti2004, Pettini2004, Erb2006b, Kewley2019, Curti2020, Gillman2021}).
At $z\sim0$, solar-metallicity galaxies typically exhibit $\log({\nii/H\alpha})\approx -0.40$, but the ratio decreases towards high redshift as metallicities decline. 
It may also be enhanced by harder ionising spectra or AGN activity.

To assess whether an AGN-like contribution could bias the adopted \nii/H$\alpha$ ratio, we measure the \oiii$\lambda$5007/H$\beta$ ratio where available, obtaining a median $\log{(\oiii\lambda 5007/{\rm H}\beta)} = 0.20$ at a median stellar mass $\log({M_{\star}/M_{\odot})} = 10.93$.
In the local Mass-Excitation (MEx) diagram \cite{Juneau2011}, such a value would fall in the AGN-dominated regime. 
However, once redshift evolution is taken into account (e.g. \citealt{Juneau2011, Coil2015, Strom2017}), this ratio is fully consistent with the star-forming locus at 
$z\sim2-3$.
Given the intrinsic 0.17 dex scatter \citep{Strom2017}, $\approx75\%$ of our galaxies are likely dominated by star formation.
We therefore assume that our sample follows this population-averaged, star-forming behaviour when estimating the \nii/H$\alpha$ correction.
Deciphering each contribution (and broad line contribution) is beyond the scope of this paper.

For 28 galaxies with grating spectra, we measure a median \nii$\lambda$6584/H$\alpha$ ratio of 0.46 (i.e. $\log{(\nii\lambda 6584/\rm H{\alpha})}$ = -0.34) and a ratio of 0.74 between the sum of both \nii\, lines and H$\alpha$\footnote{If we restrict samples with $S/N>3$ for all lines, the median value of the line ratio (\nii$\lambda\lambda$6548+6583/H$\alpha$) becomes smaller by $\approx$20\% with a higher median redshift ($\langle z \rangle$ = 3.14).}.
We adopt the latter to correct the blended PRISM flux, multiplying \texttt{line\_ha\_nii} by 0.57.
Figure~\ref{fig:n2ratio} shows $\log{(\nii\lambda 6584/\rm H{\alpha}})$ versus stellar mass for these 28 galaxies.

For comparison, \citet{Sanders2021} reported a \nii/H$\alpha$ ratio that is approximately 0.29 dex lower for star-forming galaxies at $z\approx 2.3$. 
Several factors may contribute to this difference, most notably the differing sample selections. The galaxies studied by \citet{Sanders2021} have, on average, stellar masses that are $\approx$0.35 dex lower than those in the \ecogal\ sample and lie at slightly lower redshifts ($\langle z\rangle \approx 2.2$ compared with $\langle z\rangle \approx 2.8-3.5$, depending on the selected subsample). In addition, their sample is selected from HST/F160W detections, whereas the \ecogal\ prior catalogue is based on detections in the longer-wavelength JWST filters (Sect.~\ref{sec:catalog}), making it more sensitive to massive, dust-obscured galaxies. 
In this regard, \citet{Shapley2020} measured \nii/H$\alpha$ ratios for four ALMA-selected galaxies at $z = 1.41-2.54$ with stellar masses in the range $10.36\leq \log(M_{\star}/M_{\odot})\leq10.91$. 
Their median line ratio of $-0.42$ (in log) is more consistent with the median value measured for the \ecogal\ sample, supporting the interpretation that sample selection, particularly towards more massive and dust-obscured galaxies, contributes to the observed offset relative to \citet{Sanders2021}.

Interpreting the \nii/H$\alpha$ ratio as a metallicity indicator through the N2 calibration (e.g. \citealt{Pettini2004, Bian2018}), the observed offset of $\approx 0.3$ dex is broadly consistent with the metallicity difference expected from the mass-metallicity relation at $z \approx 2-3$ (e.g. \citealt{Sanders2021}). This interpretation is further supported by previous studies showing that dust-selected galaxies tend to exhibit relatively high metallicities inferred from N2 diagnostics (e.g. \citealt{Casey2017, Shapley2020, Gillman2026}).

Nevertheless, caution is warranted when interpreting \nii/H$\alpha$ as a direct metallicity tracer. The scatter in the mass–metallicity relation is substantial at the high-mass, high-SFR end, particularly for dust-selected galaxies, and the N2 ratio is also sensitive to variations in ISM conditions, including shocks and AGN activity (e.g. \citealt{Kewley2001}). 
Furthermore, systematic uncertainties associated with the N2 calibration itself remain significant (e.g.  \citealt{Pettini2004, Sanders2018}). A more comprehensive investigation combining multiple emission-line diagnostics would therefore be required to determine the origin of the elevated \nii/H$\alpha$ ratios observed in the \ecogal\ sample, but such an analysis is beyond the scope of this paper.

Finally, we accounted for slit losses using the DJA broadband photometry by linearly interpolating between the two filters bracketing the H$\alpha$ and H$\beta$ wavelengths. 
After applying this correction, our final working sample consists of 28 galaxies.

\subsubsection{Balmer decrement}

We corrected the H$\alpha$ flux for dust attenuation using the Balmer decrement.
The Balmer decrement quantifies the excess of the observed H$\alpha$/H$\beta$ ratio above the intrinsic ratio expected in the absence of dust.
For this intrinsic value, we adopted the Case B \citep{Baker1938} recombination ratio of 2.86, corresponding to an electron temperature of $T_e = 10^4$ K and an electron density of 100 cm$^{-3}$ \citep{Osterbrock1989}. 
This is the standard assumption for star-forming galaxies, where Lyman-series photons are scattered (optically thick) and ultimately converted into Balmer-series photons, higher-order recombination lines, or two-photon continuum emission.

We translated the observed H$\alpha$ and H$\beta$ ratio to the colour excess (of nebular emission) based on the following equation:
\begin{align}\label{eq:ebv}
    E(B-V)_{\rm neb} &= \frac{2.5}{k(H\beta) - k(H\alpha)} \times \log{_{10}{\left[\frac{({\rm H\alpha/H\beta})_{\rm obs}}{2.86}\right]}} \nonumber\\
    &\simeq 1.97 \times \log{_{10}{\left[\frac{({\rm H\alpha/H\beta})_{\rm obs}}{2.86}\right]}}
,\end{align}
where we take the \citet{Calzetti2000} extinction curve.
The excess will be $\approx$10\% lower than one would get when assuming the \citet{Cardelli1989} extinction curve for \hii\, regions.

We used the same dust extinction curve to infer the magnitude of dust extinction for H$\alpha$ flux.
The inferred $A_{{\rm H}\alpha}$ was calculated with
\begin{equation}
    A_{\rm H\alpha} = 3.33 \times E(B-V)_{\rm neb}
.\end{equation}

The left panel of Fig.~\ref{fig:sfr} shows the comparison between inferred colour excess ($E(B-V)_{\rm neb}$) for the nebular component from the H$\alpha$ and H$\beta$ line ratios and the stellar component from $A_{\rm V}$ based on the \eazypy\ result, which uses the \citet{Calzetti2000} law.
To guide the eye, we also show two lines for a 1:1 relation and the relation $E(B-V)_{\rm neb} = E(B-V)_{\rm stars}/0.44$ from \citet{Calzetti2000}.

The results show that, for most galaxies, the nebular emission is more highly reddened than the stellar continuum. 
This extends the trend previously observed at $z\sim2$ galaxies (e.g. \citealt{Reddy2015}), out to $z\sim6$.
The degree of excess nebular reddening is consistent with a picture in which galaxies with higher SFRs host more dust-enriched H II regions, leading to stronger attenuation of their nebular lines (e.g. \citealt{Calzetti1994, Reddy2015}). 

We also identify six cases in which the observed H$\alpha$/H$\beta$ ratio falls below the theoretical Case B limit (giving blue colour `excess').
Three of these have weak H$\beta$ flux with S/N of $<2$, resulting in large uncertainties in the measured flux ratios. 
The remaining three cases are observed in the NIRSpec/PRISM.
Two are lower-mass systems ($<10^{10.5}\, M_{\odot}$), for which our assumed correction for the \nii\ contribution may be overestimated. The last case likely has a significant AGN contribution, as indicated by its high $\log{(\oiii/{\rm H}\beta)} \approx0.7$).
For all six galaxies, the measured H$\alpha$/H$\beta$ (and therefore $E(B-V)_{\rm neb}$) should be treated as a lower limit when applying dust attenuation corrections to the H$\alpha$ flux. 
Another mechanism that can suppress  H$\beta$ relative to H$\alpha$ is Balmer absorption from older stellar populations -- plausible here given their high stellar masses, close to, or exceeding $10^{11}\, M_{\odot}$.
These sources are marked as open triangles in the right panel of Fig~\ref{fig:sfr}.

\begin{figure*}[thb]
\centering
\includegraphics[width=0.98\textwidth]{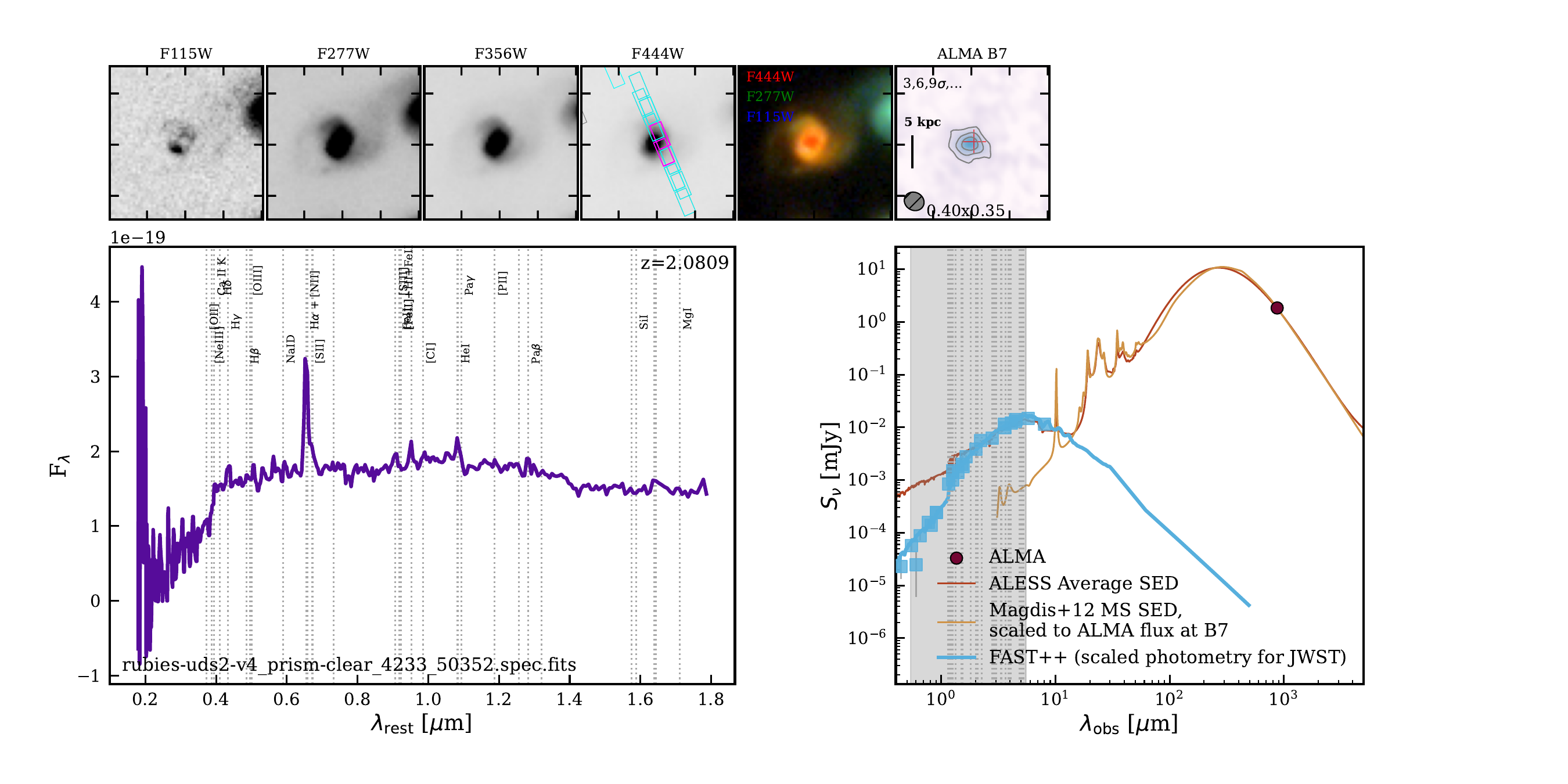}
\caption{Example of the \ecogal\, data product for a galaxy at $z=2.08$. Top panels: JWST/NIRCam (F115W, F2772W, F356W, and F444W) and ALMA Band 7 images. The galaxy has NIRSpec PRISM spectra, with a clear H$\alpha$ detection (bottom left panel). On the right, we show the best-fit SED from \textsc{fast++} using optical/NIR photometry, along with MIR/FIR SED templates \citet{Magdis2012} and \citet{daCunha2015} scaled to the ALMA flux. A DJA post accompanying this paper provides instructions to retrieve similar plots for spectroscopically confirmed, ALMA-detected sources.\label{fig:showcase}}
\end{figure*}

We compare the corrected SFR from the Balmer decrement with SFRs inferred from the dust continuum and \eazypy \,fit.
For the FIR dust SED, 
we assume a greybody model with a dust emissivity index $\beta=2.03$, following \citet{Berta2016}, and a dust temperature that evolves with redshift according to \citet{Schreiber2018b}.
The right panel of Fig.~\ref{fig:sfr} presents the comparison between dust-corrected H$\alpha$ SFRs, dust-based SFRs, and those derived from \eazypy\ fits.

The dust-based SFRs are, on average, higher than the dust-corrected H$\alpha$ SFRs, with median offsets of 0.18 dex for the grating sample and 0.49 dex for the prism sample. Several factors may contribute to these differences. First, the two tracers probe star formation over different timescales: H$\alpha$ is sensitive to recent star formation on timescales of $<10$ Myr, whereas dust-based tracers average the star formation activity over longer periods ($\sim$100 Myr). Second, our dust-based SFR estimates rely on simplified assumptions of fixed dust temperature and emissivity. For example, increasing the assumed dust temperature by 3 K would increase the inferred SFR by approximately 50\% and vice versa.
In addition, we applied a fixed correction for the \nii\ contribution to the prism spectra, which may not fully capture the diversity of line ratios among galaxies. 
Finally, a subset of heavily dust-obscured galaxies may still have underestimated attenuation corrections in both the Balmer decrement and SED-derived SFRs. While the first three factors primarily affect the comparison between the dust- and H$\alpha$-based SFR estimates, underestimated dust attenuation in heavily obscured systems may also lead to systematically underestimated SED-derived SFRs. This could explain the higher fraction of galaxies with low specific SFRs compared to UVJ-selected samples.

A comprehensive assessment of the relative importance of these effects would require additional constraints on the dust properties, emission-line diagnostics, and star formation histories. In this work, our aim is to establish population-wide trends and provide measurements that can be further refined using the tools and methodologies best suited to individual scientific applications.

Finally, the colours in the left panel of Fig.~\ref{fig:sfr} reflect the sSFR values derived from dust-based SFRs.
Some high-sSFR sources appear to have relatively low $E(B-V)_{\rm neb}$ and similarly low stellar $E(B-V)_{\rm star}$ regimes in contrast with the trend reported by \citet{Reddy2015} (such as in their Fig. 16), which shows increasing nebular colour excess with higher sSFR. 
A deeper understanding of these objects requires higher spatial-resolution observations, which would allow for a detailed investigation of the dust–nebula–star geometry.


\section{Scientific outlook}\label{sec:science}

We anticipate that the \ecogal\ product will offer a wide range of scientific use cases to deepen our current knowledge of galaxy evolution.
Figure~\ref{fig:showcase} is an example of a galaxy at $z=2.08$ that summarises the existing data (JWST photometry and spectra, ALMA cutout, as well as \textsc{fast++} fitting results) to an ALMA detection in Band 7.
Here, we briefly outline several directions to explore.

\subsection{Dust-based gas masses}\label{sec:mugas}
The dust continuum properties of galaxies beyond $z>3$ remain poorly characterised. The \ecogal\ data products provide dust continuum measurements for a large sample of galaxies that, when combined with the extensive spectroscopic information, provide the basis for estimating gas masses using standard dust-based calibrations. 
These measurements facilitate studies of gas fractions, depletion times, and gas scaling relations over a broad range of galaxy properties.
\ecogal\ complements previous surveys such as \acosmos\ and \agds\ by extending the coverage to the UDS field, incorporating JWST photometric and spectroscopic observations, and substantially increasing the number of galaxies with dust continuum measurements at $z>3$  (Lee et al. 2027, in prep.).

An interesting extension of this effort is to explore additional parameter spaces, such as the galaxy environment. 
Existing results on the gas and dust content of galaxies in distant overdense regions -- specifically, whether they are more or less gas-rich than their field counterparts -- remain inconclusive. (e.g. \citealt{minju2017,minju2021b, Hayashi2018, Darvish2018, Williams2022, Alberts2022, Gururajan2025}).

The rightmost panel of Fig.~\ref{fig:galparams} shows the distribution of galaxy overdensities, quantified using the tenth-nearest neighbour estimator, as a function of stellar mass.
For this demonstration, we restricted the sample to spectroscopically confirmed galaxies within the ALMA-covered regions, constructing true 3D densities.
We used \textsc{galcluster}\footnote{\url{https://github.com/Nikolaj-B-Sillassen/galcluster}}(\citealt{Sillassen2022}) for each legacy field, computing distances in redshift space relative to the central redshift of the catalogue volume at $z=1.5$.
The significance of overdensity was derived by accounting for the log-normal distribution of overdensities in each field.

Galaxies with ALMA detection (and spec-$z$) are colour-coded by gas fraction ($M_{\rm gas}/M_{\star}$). 
Gas fractions are estimated from the ALMA dust continuum using a greybody model with a redshift-dependent dust temperature \citep{Schreiber2018c}, and a fixed dust-to-gas mass ratio of 92.
As expected from the established scaling relations, the gas fraction decreases with increasing stellar mass. 
In contrast, no clear dependence of gas fraction on environment is visible, at least from a qualitative inspection.
Although environmental influence is likely a secondary (or even tertiary) factor in regulating the cold molecular gas and dust content of galaxies, \ecogal\ will allow for a significantly more detailed and quantitative investigation of these trends.

\subsection{Obscured star formation activities}

\begin{figure}[t]
\centering
\includegraphics[width=0.48\textwidth]{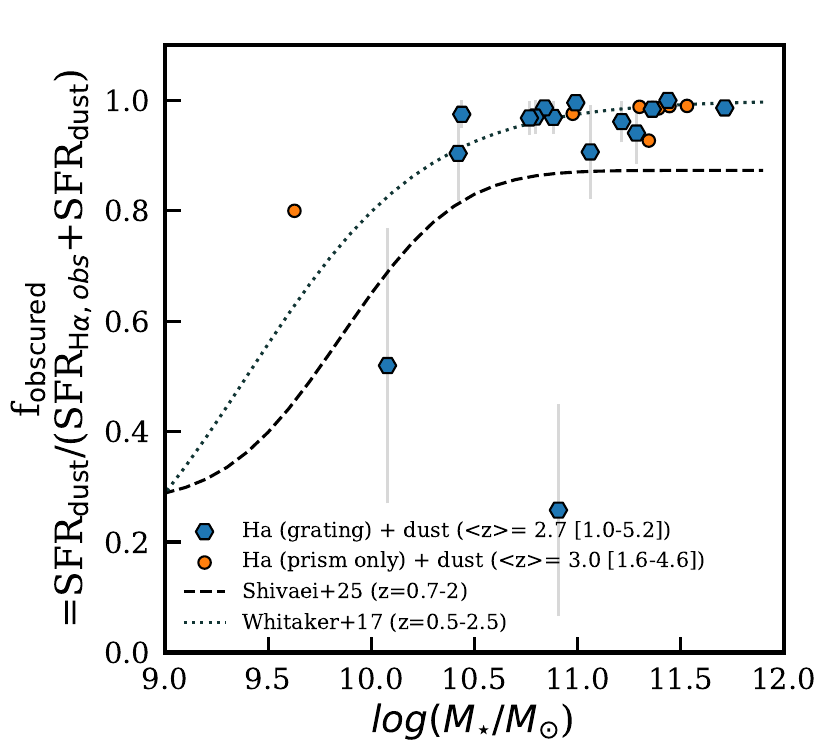}
\caption{Obscured star formation (1-SFR(attenuated, H$\alpha$)/SFR(total) as a function of stellar mass for galaxies with JWST spectra and ALMA detection. Empirical relations for $z\lesssim2.5$ galaxies are shown from \citet{Whitaker2017} and \citet{Shivaei2024}. ALMA-detected galaxies occupy the highest dust-obscured fractions, broadly consistent with \citet{Whitaker2017}, which was based on \spitzer/MIPS.\label{fig:fobsc}}
\end{figure}

\begin{figure*}[th]
\centering
\includegraphics[width=0.98\textwidth]{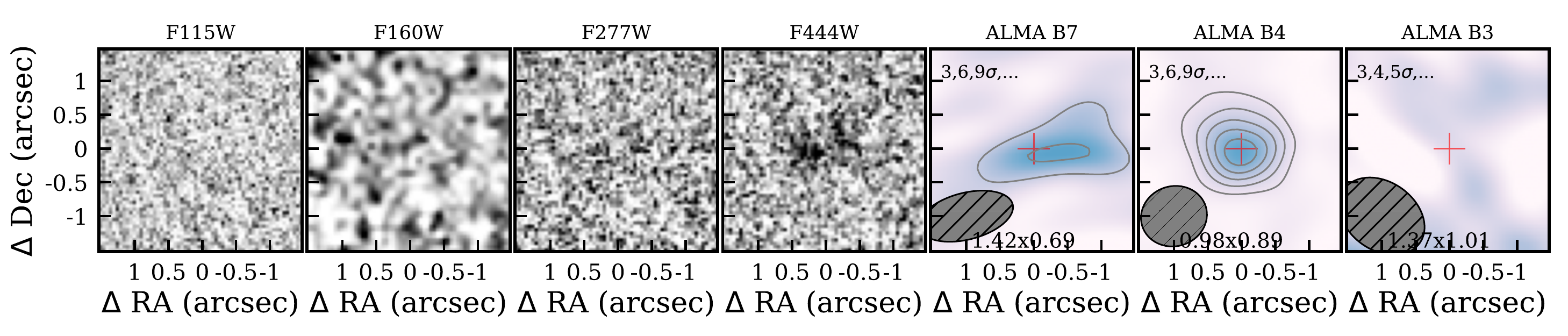}
\caption{Example of an ALMA-detected galaxy with no counterpart in the existing F444W catalogue, likely representing an optically dark source. The panels show cutouts in HST/F160W and JWST/NIRCam F115W, F277W, and F444W filters, along with ALMA Bands 3, 4, and 7\label{fig:optdark}.}
\end{figure*}

\begin{figure}[th]
\centering
\includegraphics[width=0.48\textwidth]{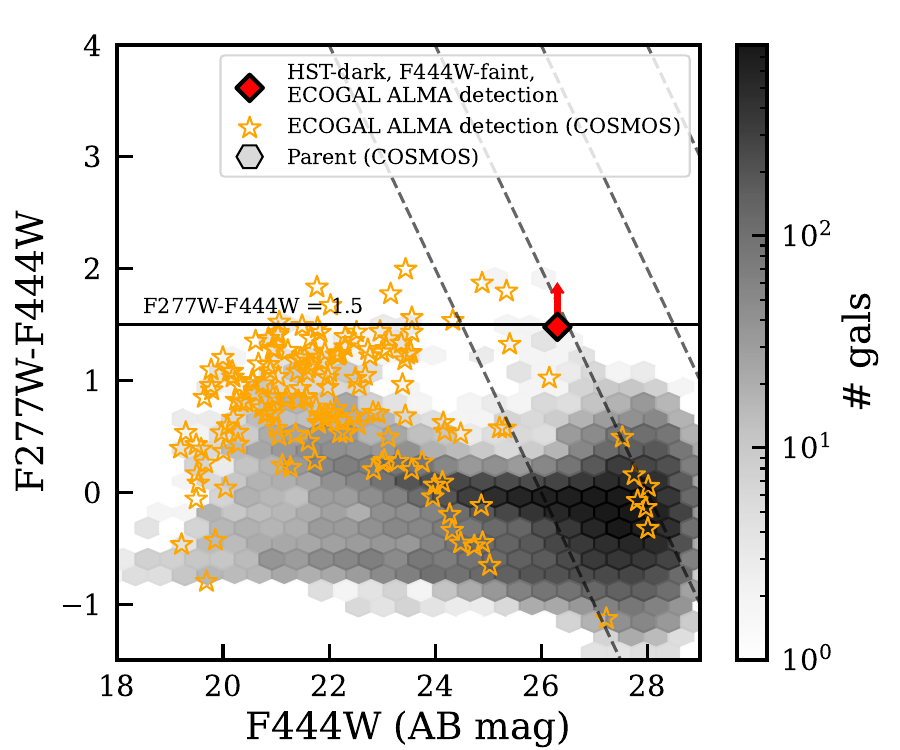}
\caption{Colour–magnitude diagram of F277W–F444W versus F444W for the COSMOS field, where the parent sample is shown in hexagonal bins coloured by the number of galaxies. ALMA-detected sources with optical/NIR counterparts are marked as open orange stars.
Only sources with $S/N >3$ in both F277W and F444W are shown. 
Dashed lines indicate fixed F277W magnitude thresholds of 26, 28, 30, and 32 mag (left to right). F277W-F444W colour cut follows \citet{Barro2024}, designed to select massive galaxies with red, dusty, or quiescent SEDs, as well as possible high-equivalent-width emission-line galaxies at $z\gtrsim5$. The highlighted source (red diamond, Fig.~\ref{fig:optdark}) is recently spectroscopically confirmed at $z=6.6$ \citep{Bing2026}.\label{fig:color}}
\end{figure}

Obscured and unobscured star formation provide complementary views of galaxy evolution and are key to understanding how galaxies build up their stellar mass across cosmic time (e.g. \citealt{Magnelli2011, Magnelli2013, Madau2014, Wang2019b, Zavala2021, Algera2023}).
By combining multiple star formation tracers available in the \ecogal\ data products, namely H$\alpha$ and dust continuum emission, we can investigate the obscured fraction of star formation in individual galaxies as a function of stellar mass at $z > 2.5$ and compare it across different galaxy populations.

Building on the measurements described in Sect.~\ref{sec:sfr}, we showcase the obscured star formation for those with ALMA detection and the JWST spectra. 
The left panel in Fig.~\ref{fig:fobsc} shows the obscured star formation, defined using
\begin{equation}
    f_{\rm obscured} = 1-\frac{SFR_{{\rm H}\alpha,{\rm observed}}}{SFR_{\rm total}}
,\end{equation}
where $SFR_{{\rm H}\alpha, {\rm observed}}$ is derived from the observed (slit-loss and \nii\ contribution corrected) H$\alpha$ flux but not corrected for the Balmer decrement.  SFR$_{\rm total}$ is the sum of dust-based SFR and $SFR_{{\rm H}\alpha, {\rm observed}}$.
For comparison, we include the empirical relations from \citet{Whitaker2017} and \citet{Shivaei2024}, which apply to galaxies $z\lesssim2.5$.

As expected, the ALMA-detected galaxies predominantly lie at high obscured fractions for their stellar masses ($f_{\rm obsc} > 0.95$) above the empirical trend of \citet{Shivaei2024} but more consistent with \citet{Whitaker2017}. 
The latter is based on the \spitzer/MIPS detection, which may be more aligned with ALMA-detected samples.
A natural next step is to compare these results with ALMA non-detections, and to trace how the obscured fraction evolves with redshift. 
The combination of JWST and ALMA data provided by \ecogal\ enables exactly this type of investigation.

\subsection{Optically dark sources}\label{sec:optdark}

Historically, galaxies selected via dust emission in the submm/mm (submillimetre galaxies (SMGs; \citealt{Blain2002}) and dusty star-forming galaxies (DSFGs; \citealt{Casey2014}) often show faint or undetectable optical/NIR emission, making it difficult to characterise their physical properties \citep[e.g.][]{Aguirre2013, Lang2019, Gillman2023, Gillman2024}.
Galaxies selected by red optical/NIR colours (e.g. H-dropout; \citealt{Caputi2012, Stefanon2015, Wang2019}) provide a complementary window, particularly at $z>3$, typically identifying less extreme but still dust-obscured systems.
Before JWST, these selections relied on clear detections in IRAC channels ([3.6] or [4.5] $\leq 24.5$ mag) and very faint fluxes in $\hst/F160W$ ($\gtrsim 26-27$ mag).
JWST now enables the identification of galaxies much fainter in the F444W or F356W \citep{Perez-Gonzalez2023, Barrufet2023}, opening new parameter space for dusty, high-redshift populations. Understanding how these systems connect to classical submm/mm–selected samples is essential for mapping the full obscured star formation history.

Some galaxies in the ECOGAL blind catalogue show no optical–NIR counterparts or are detected only marginally in F444W (Sect.~\ref{sec:blind}).
If real, such sources may represent extremely dust-obscured galaxies and/or systems at $z\gtrsim3$, where red colours may trace the Balmer break.
Constraining the number densities of these galaxies is therefore crucial for understanding obscured star formation at $z>3$, and the emergence of dust in the earliest Universe.

To showcase the unique parameter space accessible with \ecogal, we highlight one example source.
Figure~\ref{fig:optdark} presents the JWST and ALMA cutouts: 
The object is clearly detected in Band 4 and shows a tentative signal in Band 7, though below the blind-detection threshold (peak $S/N<6.6$).
In contrast, the HST and JWST imaging reveals only a faint detection in the longest-wavelength NIRCam filter (F444W).

Because this source was not included as a detection in the DJA parent catalogue, we performed aperture photometry to measure its F277W-F444W colour and F444W magnitude. 
A circular aperture of 1-arcsec diameter was chosen based on the curve-of-growth analysis, where the curve reaches a plateau. 
This aperture is more than twice as large as typical choices in the literature for colour selection (e.g. \citealt{Labbe2023, Barro2024}) and the adopted aperture size for the DJA catalogue, reflecting the source’s extended morphology.

Figure~\ref{fig:color} shows the colour-magnitude diagram for the \ecogal\ parent galaxies in the COSMOS field with ALMA coverage.
The highlighted source lies in a sparsely populated region with F444W$>$26 mag, but a very red colour ($F277W-F444W\approx1.5$).
In contrast, most ALMA-detected galaxies with optical counterparts (orange stars) have significantly brighter F444W magnitudes.
Previous work has explored similar colour selections ($F277W-F444W > 1.5$, $F444W<28$ mag) in the CEERS field, finding a median photometric redshift of $z=6.9^{+1.0}_{-1.6}$ \citep{Barro2024}.
However, those samples appeared point-like in all NIRCam bands. This suggests that ALMA may be selecting more massive and dust-rich systems at $z>6$, whose optical/NIR emission appears more extended.
Notably, the source we highlight has recently been spectroscopically confirmed at  $z=6.63$ (\citealt{Bing2026}).
Deeper MIR observations and targeted spectroscopy for similarly optically dark \ecogal\ sources will be crucial for uncovering their nature and for tracing obscured star formation in the early Universe.

\section{Summary}\label{sec:summary}
In this paper, we present an overview of \ecogal, an ALMA data-mining product covering three major extragalactic legacy fields (COSMOS, GOODS-S, and UDS).
\ecogal\ enhances the value of these fields by integrating publicly available \jwst\ data through the DJA framework. 
Our first public data release includes two ALMA catalogues and the associated FITS files: one catalogue based on \jwst\ prior positions and another constructed from blind detections within the ALMA coverage.
\ecogal\ provides ALMA continuum flux constraints of $\sim$130,000 sources spanning a wide range of redshifts, substantially expanding the parameter space for scientific investigation.
This resource enables a broad range of studies relevant to galaxy formation and evolution. We highlighted several key science cases, including the dust-based gas mass measurement, obscured star formation, and an optically dark source to illustrate \ecogal’s potential.
Beyond individual detections, \ecogal\ also delivers extensive spectroscopic redshift information and continuum maps, enabling detailed investigations of interstellar medium properties in both star-forming and quiescent galaxies through methods such as stacking analyses.
This first release focuses on dust continuum detections, but we plan to extend \ecogal\ to additional fields and to include spectral data cubes as resources allow. 
We hope that \ecogal\ will be widely used by the community and will enable new and inspiring scientific discoveries.
A companion post on the DJA webpage is released alongside this paper\footnote{\url{https://dawn-cph.github.io/dja/blog/2025/11/20/ecogal-dja-showcase/}}, detailing how to access the data and perform initial visual checks using either source IDs or sky co-ordinates. The GitHub repository for the query functionality based on the data release is publicly available\footnote{\url{https://github.com/mjastro/ecogal}}.

\section*{Data availability}

A full description of the catalogue columns is available on AWS via the following link: \url{https://s3.amazonaws.com/alma-ecogal/dr1/catalogue/ecogal_v1.1.columns.csv}. The first data release of the detection catalogue, combining prior-based and blind detections and including the \textsc{clean}-related flag (\texttt{flag\_clean}) and line-contamination flag (\texttt{flag\_line}), is also available on AWS via the following link: \url{https://s3.amazonaws.com/alma-ecogal/dr1/catalogue/ecogal_detection_combined-prior_and_blind_v1.1.csv}. The AWS-hosted data products (within DJA) are intended to provide a flexible and updateable version of the catalogue. For a stable, citable version of the catalogues without the additional flags, we provide the prior-based and blind-detection catalogues through Zenodo: \url{https://doi.org/10.5281/zenodo.17640217}.

\begin{acknowledgements}
The authors thank the anonymous referee for their constructive report.
This project has received funding from the European Union’s Horizon 2020 research and innovation programme under the Marie Skłodowska-
Curie grant agreement No 101107795.
The data products presented herein were retrieved from the Dawn JWST Archive (DJA). 
DJA is an initiative of the Cosmic Dawn Center (DAWN), which is funded by the Danish National Research Foundation under grant DNRF140.
This work was supported by the  DeiC National
HPC (DeiC-DTU-N2-2024057).
BG and CH acknowledge support from Villum Fonden under project grant 53120.
This paper makes use of the following ALMA data: ADS/JAO.ALMA\#2011.0.00097.S,
\#2011.0.00115.S, \#2011.0.00124.S, \#2011.0.00648.S, \#2011.0.00716.S,
\#2012.1.00033.S, \#2012.1.00245.S, \#2012.1.00307.S, \#2012.1.00978.S,
\#2012.1.00983.S, \#2013.1.00034.S, \#2013.1.00092.S, \#2013.1.00118.S,
\#2013.1.00139.S, \#2013.1.00146.S, \#2013.1.00151.S, \#2013.1.00171.S,
\#2013.1.00205.S, \#2013.1.00208.S, \#2013.1.00250.S, \#2013.1.00276.S,
\#2013.1.00470.S, \#2013.1.00566.S, \#2013.1.00742.S, \#2013.1.00781.S,
\#2013.1.00786.S, \#2013.1.00836.S, \#2013.1.00884.S, \#2013.1.00914.S,
\#2013.1.01271.S, \#2013.1.01292.S, \#2015.1.00026.S, \#2015.1.00039.S,
\#2015.1.00040.S, \#2015.1.00055.S, \#2015.1.00137.S, \#2015.1.00207.S,
\#2015.1.00228.S, \#2015.1.00242.S, \#2015.1.00260.S, \#2015.1.00379.S,
\#2015.1.00388.S, \#2015.1.00442.S, \#2015.1.00456.S, \#2015.1.00568.S,
\#2015.1.00664.S, \#2015.1.00695.S, \#2015.1.00704.S, \#2015.1.00821.S,
\#2015.1.00861.S, \#2015.1.00870.S, \#2015.1.00907.S, \#2015.1.00948.S,
\#2015.1.01074.S, \#2015.1.01096.S, \#2015.1.01105.S, \#2015.1.01111.S,
\#2015.1.01129.S, \#2015.1.01205.S, \#2015.1.01212.S, \#2015.1.01222.S,
\#2015.1.01379.S, \#2015.1.01447.S, \#2015.1.01495.S, \#2015.1.01528.S,
\#2015.A.00009.S, \#2015.A.00026.S, \#2016.1.00048.S, \#2016.1.00142.S,
\#2016.1.00171.S, \#2016.1.00279.S, \#2016.1.00434.S, \#2016.1.00463.S,
\#2016.1.00478.S, \#2016.1.00564.S, \#2016.1.00567.S, \#2016.1.00624.S,
\#2016.1.00646.S, \#2016.1.00721.S, \#2016.1.00726.S, \#2016.1.00735.S,
\#2016.1.00776.S, \#2016.1.00790.S, \#2016.1.00967.S, \#2016.1.00990.S,
\#2016.1.01001.S, \#2016.1.01012.S, \#2016.1.01040.S, \#2016.1.01079.S,
\#2016.1.01184.S, \#2016.1.01208.S, \#2016.1.01240.S, \#2016.1.01262.S,
\#2016.1.01454.S, \#2016.1.01604.S, \#2017.1.00001.S, \#2017.1.00046.S,
\#2017.1.00190.S, \#2017.1.00270.S, \#2017.1.00326.S, \#2017.1.00373.S,
\#2017.1.00413.S, \#2017.1.00428.L, \#2017.1.00486.S, \#2017.1.00562.S,
\#2017.1.00604.S, \#2017.1.00893.S, \#2017.1.01027.S, \#2017.1.01163.S,
\#2017.1.01176.S, \#2017.1.01276.S, \#2017.1.01347.S, \#2017.1.01471.S,
\#2017.1.01492.S, \#2017.1.01512.S, \#2017.1.01618.S, \#2017.1.01659.S,
\#2017.A.00013.S, \#2018.1.00164.S, \#2018.1.00216.S, \#2018.1.00429.S,
\#2018.1.00478.S, \#2018.1.00543.S, \#2018.1.00567.S, \#2018.1.00570.S,
\#2018.1.00635.S, \#2018.1.00992.S, \#2018.1.01044.S, \#2018.1.01079.S,
\#2018.1.01128.S, \#2018.1.01225.S, \#2018.1.01281.S, \#2018.1.01359.S,
\#2018.1.01521.S, \#2018.1.01551.S, \#2018.1.01739.S, \#2018.1.01824.S,
\#2018.1.01841.S, \#2018.1.01871.S, \#2018.A.00037.S, \#2019.1.00102.S,
\#2019.1.00244.S, \#2019.1.00337.S, \#2019.1.00397.S, \#2019.1.00477.S,
\#2019.1.00678.S, \#2019.1.00702.S, \#2019.1.00900.S, \#2019.1.00909.S,
\#2019.1.01127.S, \#2019.1.01142.S, \#2019.1.01201.S, \#2019.1.01238.S,
\#2019.1.01275.S, \#2019.1.01329.S, \#2019.1.01398.S, \#2019.1.01528.S,
\#2019.1.01537.S, \#2019.1.01600.S, \#2019.1.01615.S, \#2019.1.01634.L,
\#2019.1.01722.S, \#2019.1.01832.S, \#2019.2.00143.S, \#2019.2.00218.S,
\#2019.2.00246.S, \#2021.1.00024.S, \#2021.1.00104.S, \#2021.1.00246.S,
\#2021.1.00280.L, \#2021.1.00421.S, \#2021.1.00505.S, \#2021.1.00666.S,
\#2021.1.00705.S, \#2021.1.00815.S, \#2021.1.01005.S, \#2021.1.01133.S,
\#2021.1.01159.S, \#2021.1.01188.S, \#2021.1.01221.S, \#2021.1.01291.S,
\#2021.1.01328.S, \#2021.1.01342.S, \#2021.1.01500.S, \#2021.1.01503.S,
\#2021.1.01650.S, \#2021.1.01676.S, \#2022.1.00055.S, \#2022.1.00076.S,
\#2022.1.00101.S, \#2022.1.00120.S, \#2022.1.00319.S, \#2022.1.00642.S,
\#2022.1.00764.S, \#2022.1.00863.S, \#2022.1.00884.S, \#2022.1.00955.S,
\#2022.1.01039.S, \#2022.1.01401.S, \#2022.1.01562.S, \#2022.1.01572.S,
\#2022.1.01604.S, \#2022.1.01644.S, \#2023.1.00336.S, \#2023.1.00367.S,
\#2023.1.00521.S, \#2023.1.00609.S, \#2023.1.00837.S, \#2023.1.01016.S,
\#2023.1.01052.S, \#2023.1.01066.S, \#2023.1.01269.S, \#2023.1.01292.S,
\#2023.1.01296.S, \#2023.1.01520.S, \#2023.1.01604.S, \#2023.1.01619.S,
\#2023.A.00021.S, \#2023.A.00037.S, \#2024.A.00007.S. 
ALMA is a partnership of
ESO (representing its member states), NSF (USA), and NINS
(Japan), together with NRC (Canada), MOST and ASIAA (Taiwan),
and KASI (Republic of Korea), in cooperation with the Republic
of Chile. The Joint ALMA Observatory is operated by ESO,
AUI/NRAO, and NAOJ.
This work used observations taken by the 3D-HST Treasury Programme (HST-GO-12177 and HST-GO-12328) with the NASA/ESA Hubble Space Telescope, which is operated by the Association of Universities for Research in Astronomy, Inc., under NASA contract NAS5-26555.
This work is based in part on observations made with the NASA/ESA/CSA \textit{James Webb} Space Telescope. The data were obtained from the Mikulski Archive for Space Telescopes at the Space Telescope Science Institute, which is operated by the Association of Universities for Research in Astronomy, Inc., under NASA contract NAS 5-03127 for JWST. These observations are associated with programme GTO
\#1180, 1207, 1210, 1212, 1214, 1215, 1286,
GO \#1671, 1810, 1879,  2198,  2565,  3543,  3567,
 4233,
 5224,
 6368,
DD \#6541,
 6585. 

\end{acknowledgements}

%
%
\bibliographystyle{aa} 
\bibliography{minjujournal_v2}

\begin{appendix} 
\nolinenumbers
\FloatBarrier
\section{Astrometry for 3D-\hst\ compared to DJA}

\begin{figure*}
\centering
\includegraphics[width=0.98\textwidth]{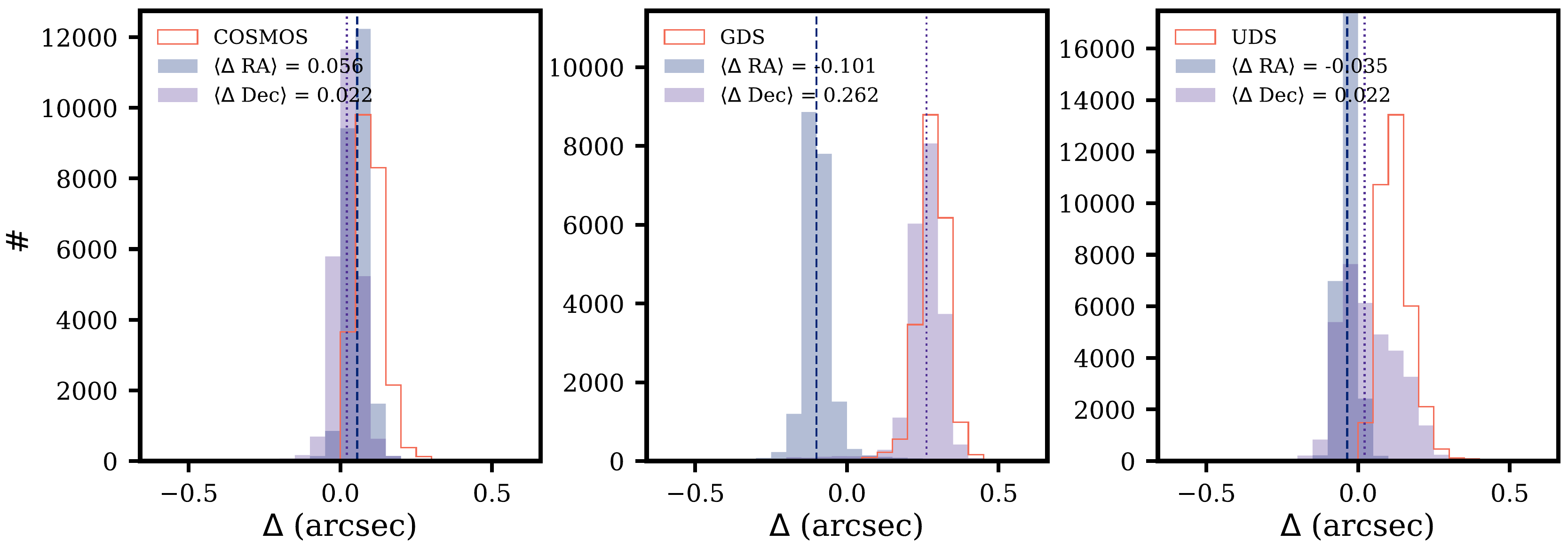}
\caption{Histogram of positional offset between \jwst\, and 3D-HST co-ordinates for COSMOS (left), GOODS-S (middle) and UDS (right). 
Horizontal axes show the offsets in right ascension (RA; navy) and declination (Dec.; purple), in arcseconds, and vertical axes show the number of galaxies per bin. The red open histogram indicates the distribution of total offset distances.
The COSMOS field shows negligible offsets, while GOODS-S exhibits a larger systematic shift, consistent with previous reports. In the UDS field, we also find an astrometric offset tail in Dec. These field-dependent offsets are corrected prior to cross-matching the ALMA positions.\label{fig:astrometry}}
\end{figure*}

We investigate the positional offset when cross-matching the sources between 3D-HST and the DJA catalogue. Figure~\ref{fig:astrometry} shows the positional offset between the sources in each field.
Galaxies in the GOODS-S present the largest offset, reconfirming the results in the literature \citet{Franco2018, Gomez-Guijarro2022}.

\section{Flag criteria for spectroscopic redshift and the robustness of photometric redshift}\label{app:specz}
In each spectroscopic catalogue, there are different redshift flags.
We considered redshift as robust when the flag is considered `secure' and `robust' (LEGA-C, KMOS3D, C3R2, PRIMUS, CANDELSz7, GOLDRUSH, JADES, UDSz).
Whenever the confidence level was available, we chose the flag that had more than 90\% confidence level (zCOSMOS, COSMOS spectroscopic catalogue, MOSDEF, VANDELS, VVDS, VUDS, VIPERS, GAMA).
Whenever the spectroscopic redshift was based on at least two line detections, we chose the corresponding flag (ZFIRE, FMOS, MUSE-Wide, DEIMOS-COSMOS).
For DJA, we used the spectroscopic redshift with grade=3.

\begin{figure*}
\centering
\includegraphics[width=0.98\textwidth]{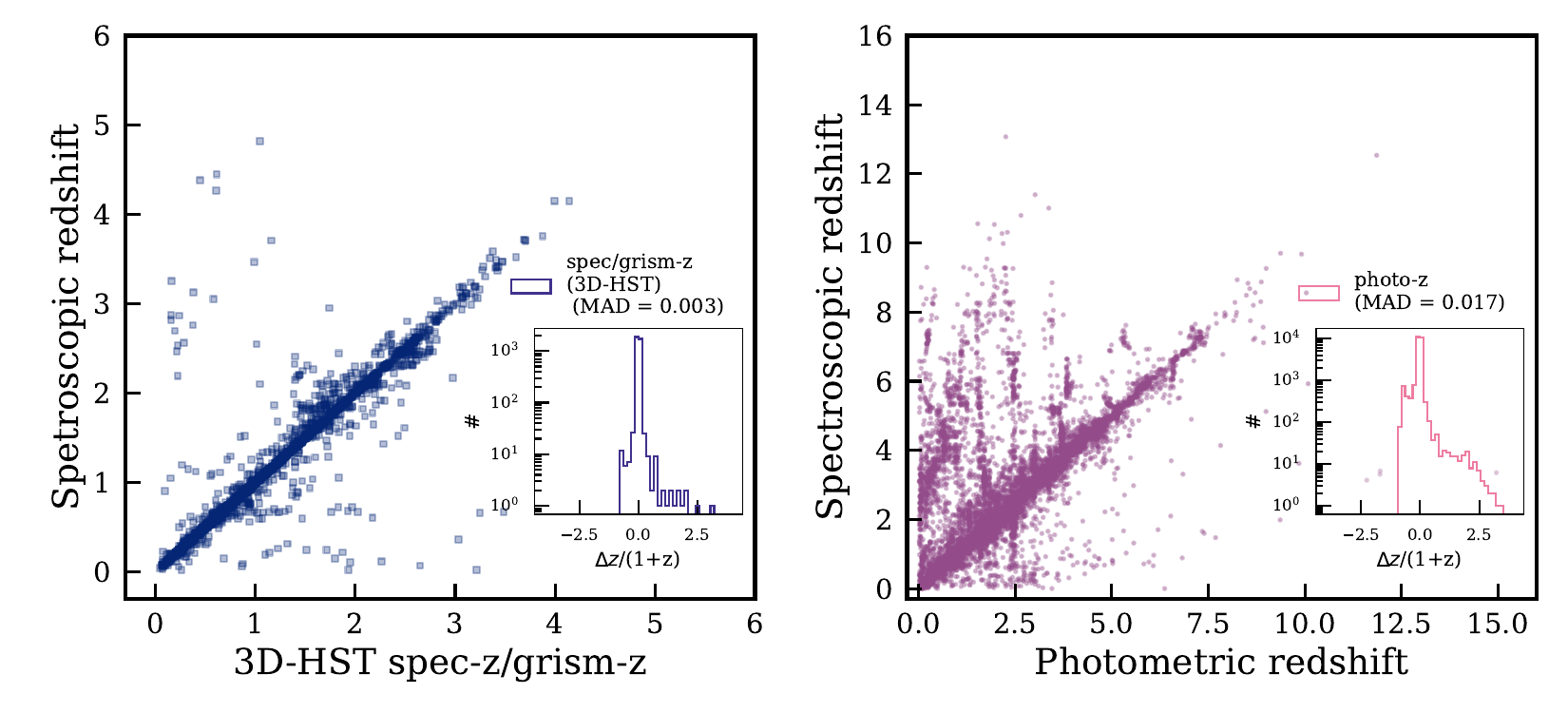}
\caption{Left: Comparison of spectroscopic redshifts from the original 3D-HST catalogue (including grism and compiled spectroscopy) with the updated spectroscopic compilation used in this work. The redshift range of the 3D-HST subset is limited to $z\lesssim6$ due to the construction of the original catalogue. Right: Comparison between spectroscopic redshifts and photometric redshifts from our \eazypy\ run. With the addition of JWST data, the accessible redshift range now extends to $z\sim15$.\label{fig:redshift}}
\end{figure*}

The newly compiled spectroscopic redshifts generally agree with the original 3D-HST redshift (Fig.~\ref{fig:redshift} left). Further, photometric redshifts from \eazypy\ show a general agreement with the compiled spectroscopic redshifts, with a MAD of 0.017 (Fig.~\ref{fig:redshift} right).

We note that some photometric redshift constraints show some strip features, which may indicate a failure of the photometric redshift from Fig.~\ref{fig:redshift}. We investigate what drives this as follows.

\begin{figure*}
\centering
\includegraphics[width=0.98\textwidth]{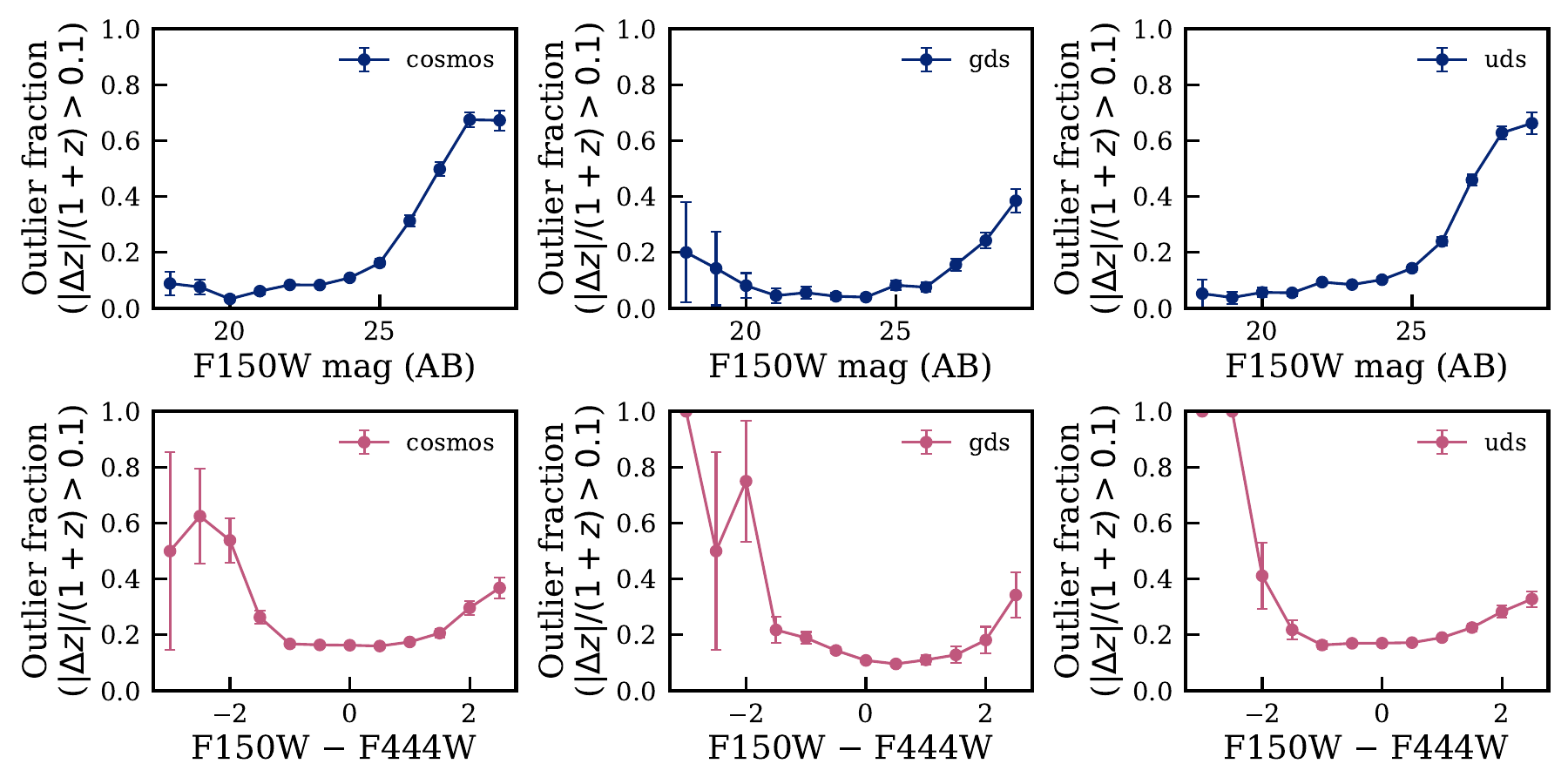}
\caption{Redshift accuracy of parent catalogue galaxies with spectroscopic redshift. Each row shows outlier fraction as a function of F150W magnitude and F150W-F444W colours, and each column shows the results for different legacy fields. Fainter galaxies and bluer and redder sources have a higher fraction of outliers.\label{fig:outliers}}
\end{figure*}

We define photometric-redshift outliers as sources with  $|\Delta z|/(1+z)>0.1$, where $\Delta z = z_{\rm spec} - z_{\rm phot, eazypy}$. This criterion follows \citet{Bezanson2016}, who evaluated the photometric-redshift performance of 3D-HST galaxies, and is more conservative than the COSMOS2020 outlier definition of \citet{Weaver2020} ($>0.15$), which follows \citet{Hildebrandt2012}.
Figure~\ref{fig:outliers} shows the dependence of the outlier fraction on $F150W$ magnitude and $F150W-F444W$ colour, indicating that the outliers are preferentially found among fainter sources and objects with extreme colours. Similarly, \citet{Bezanson2016} and \citet{Weaver2020} reported an increasing outlier fraction towards fainter observed magnitudes (e.g. JH magnitude, i-band magnitude). For example, in the COSMOS2020 catalogue, the outlier fraction increases to approximately 22\% at the faint end ($i>25$; \citealt{Weaver2020}). Restricting our sample to sources with $F150W>24$ yields an outlier fraction comparable to those reported in previous studies.

We additionally assess the outliers using the photometric-redshift probability distributions of $P(z)$ from \eazypy. 
Defining outliers as sources whose spectroscopic redshifts fall outside the 95\% confidence interval (2$\sigma$), we find an outlier fraction of approximately 8-9\% for the full sample. These sources typically have fewer valid photometric bands available for the fit, resulting in larger photometric-redshift uncertainties. 
For example, in the UDS and COSMOS fields, the median number of filters available for outliers is less than twice and three times that of the well-constrained photometric-redshift samples, respectively (12 versus 26 filters and 17 versus 52 filters). The outliers also tend to have lower S/Ns in the available bands. In particular, the UDS and COSMOS outliers show lower S/Ns in the $F125W, F140W$, and $F160W$ bands (median S/N $\approx 3-4$), which are important for identifying the Balmer break at $z\approx2-3$. The reduced photometric constraints in these bands may therefore contribute to increased uncertainties and systematic features in the photometric-redshift distribution around these redshifts.

We encourage users to inspect the released DJA photometry and, where necessary, the underlying processed images when investigating individual outliers. In some cases, apparent discrepancies may also arise from uncertainties in the spectroscopic redshift or the source cross-matching. A detailed investigation of these individual cases requires source-by-source analysis and is left to future studies.

\section{Flux measurement discrepancies and the imaging strategy}\label{app:photometry}

\begin{figure}
\centering
\includegraphics[width=0.48\textwidth]{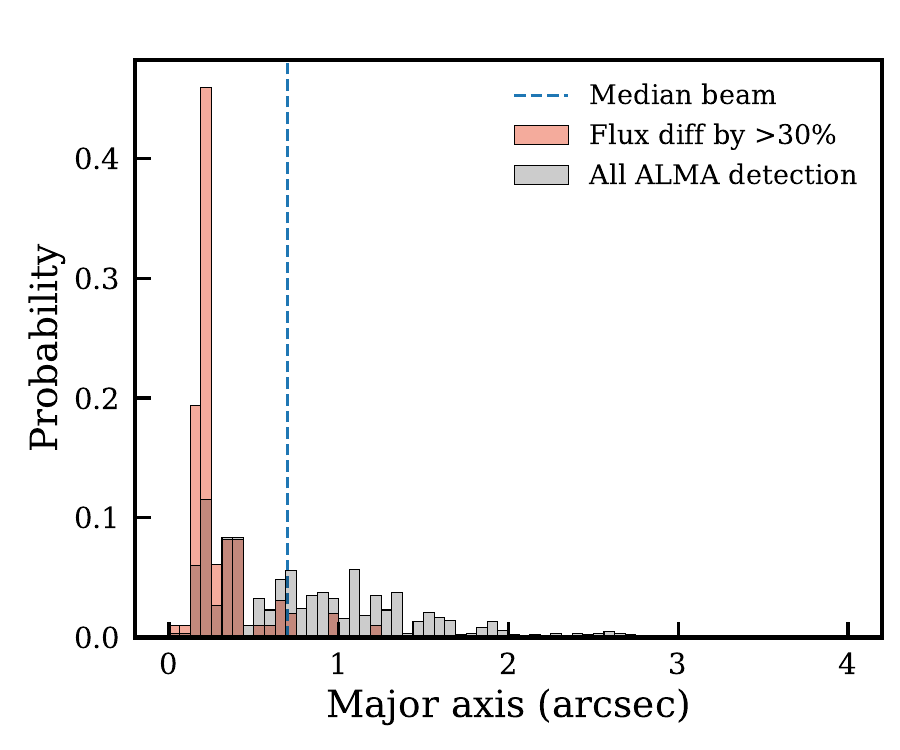}
\caption{Beam size distribution (probability density) for ALMA-detected sources, highlighting cases where flux measurements differ by more than 30\% between methods. Discrepancies tend to occur in higher spatial resolution datasets.\label{fig:largeflux_aper}}
\end{figure}

\begin{figure*}
\centering
\includegraphics[width=0.98\textwidth]{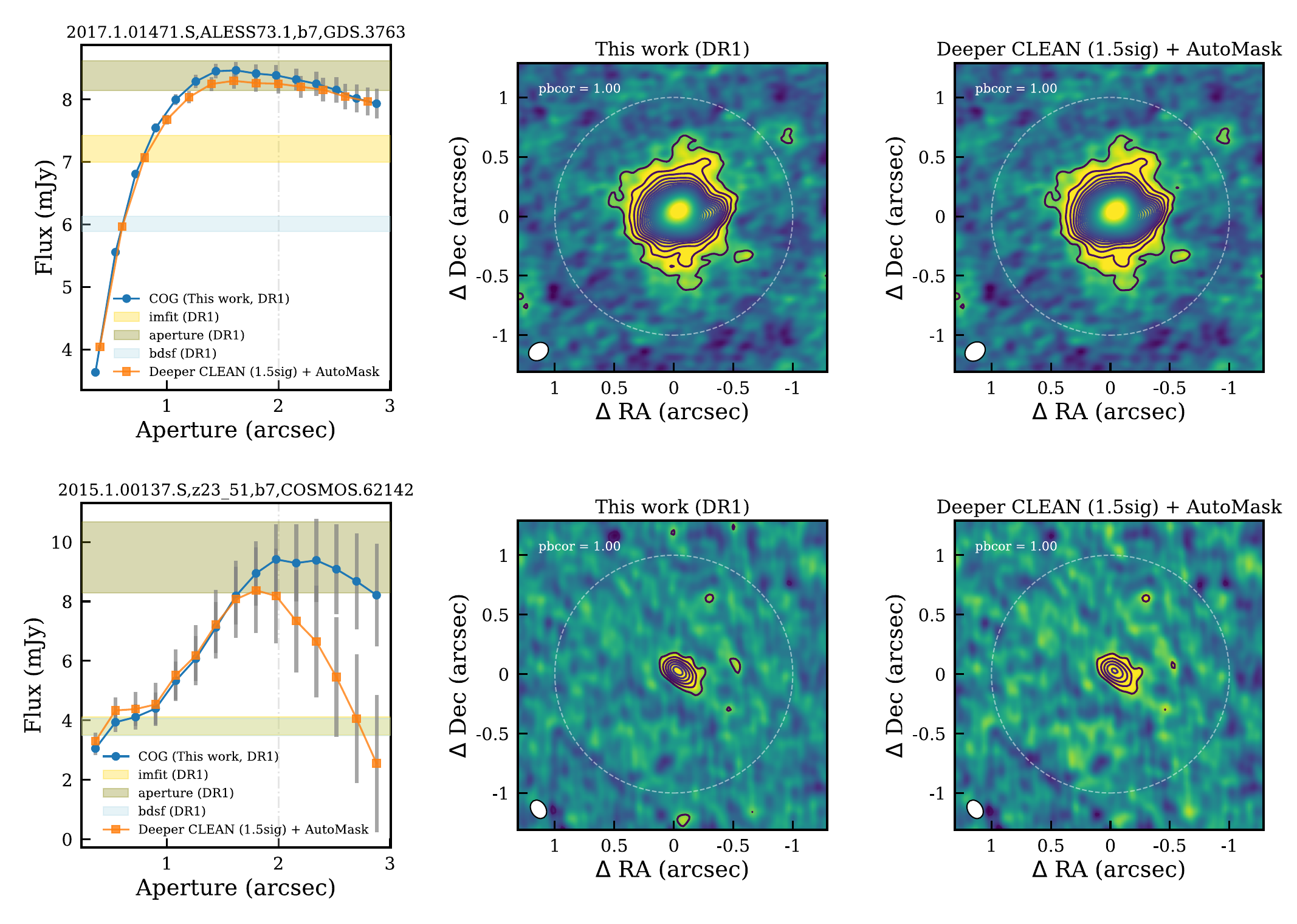}
\caption{Curves of growth for the two galaxies showing discrepant flux measurements across different methods. We plot the fluxes derived from aperture photometry, \texttt{imfit}, and \textsc{pybdsf} on the left panel for each galaxy. The dashed vertical line indicates the aperture size in diameter. The right two panels for each galaxy show the corresponding cut-out image using different \textsc{clean} parameterisation and depth, with the dashed circular aperture indicating where the aperture flux was measured. These two examples demonstrate that the flux measurements obtained with a 2-arcsec aperture are consistent across different \textsc{clean} parameterisations, while also highlighting the differences between the aperture-photometry and Gaussian-fitting flux estimates.\label{fig:flux_disagree_ok}}
\end{figure*}

\begin{figure*}
\centering
\includegraphics[width=0.98\textwidth]{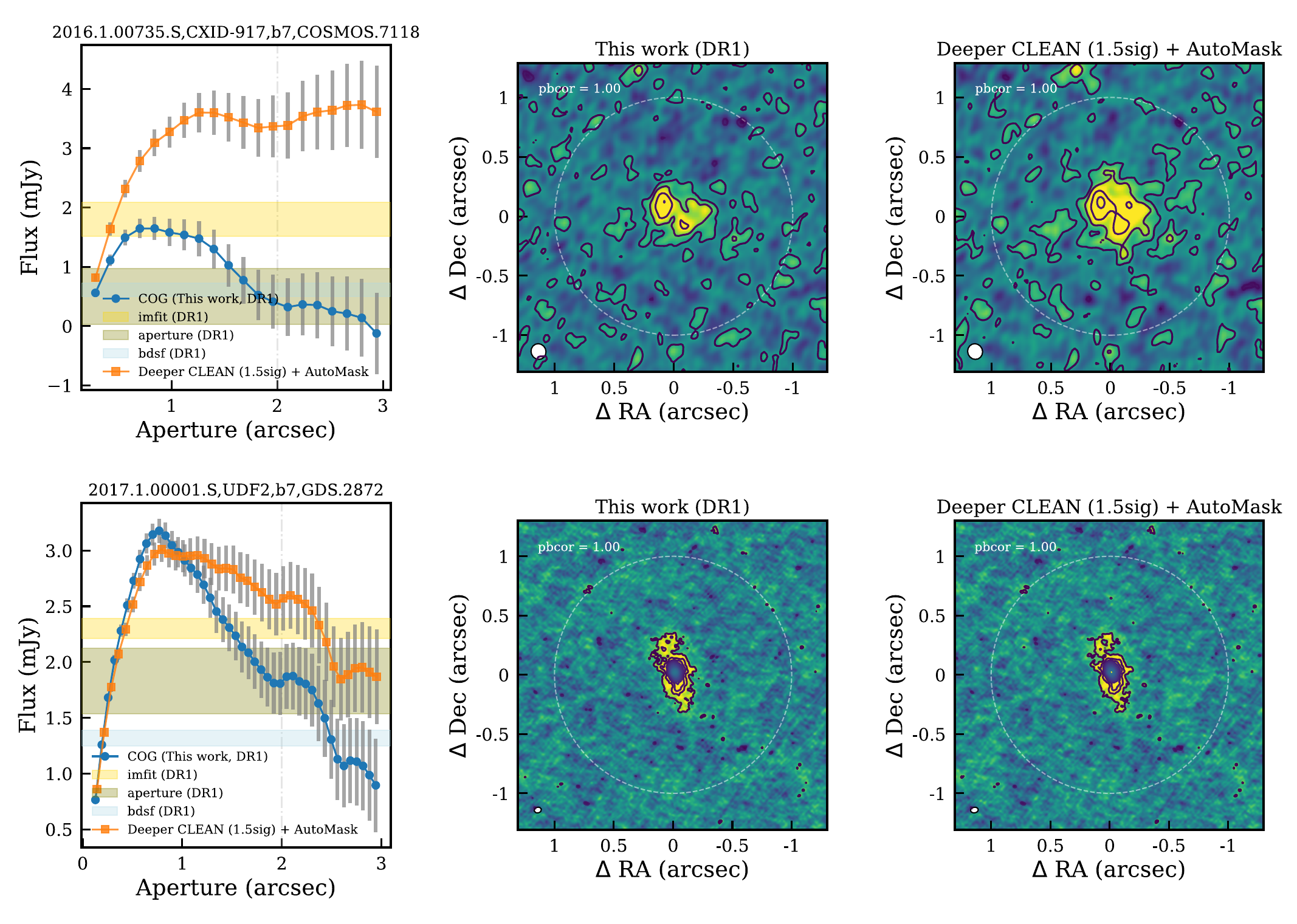}
\caption{ Same as Fig.~\ref{fig:flux_disagree_ok} but for sources whose flux measurements are inconsistent across different \textsc{clean} parameterisations.\label{fig:flux_disagree_bad}}
\end{figure*}

Figure~\ref{fig:largeflux_aper} shows the distribution of synthesised beam sizes for the detected sources, highlighting those with larger discrepancies between the two flux measurement methods. Compared to the median beam size of ($\sim 0\farcs7$), sources with discrepant measurements ($>30\%$) tend to have smaller synthesised beams, indicating that the issue is more prominent in higher-resolution datasets. 
Figures~\ref{fig:flux_disagree_ok} and~\ref{fig:flux_disagree_bad} show examples of sources with larger discrepancies between different flux extraction approaches, demonstrating that a single-component Gaussian model may not always recover the total emission and that a fixed aperture size may not be optimal for all source morphologies.

We further investigated the origin of these discrepancies and found that they may also be related to the uniform \textsc{clean} strategy adopted for the full sample. For a subset of high-resolution datasets, the flux measurements in the first data release may therefore be affected by residual imaging artefacts. To assess this effect, we re-imaged these datasets using deeper \textsc{clean}ing with adaptive automasking, following the ALMA pipeline automasking scheme by adjusting the relevant parameters (\texttt{noisethreshold}=4.25, \texttt{sidelobethreshold}=1.0, and \texttt{lownoisethreshold}=0.5) and adopting a lower \textsc{clean} threshold of $1.5\sigma$.

Figures~\ref{fig:flux_disagree_ok} and~\ref{fig:flux_disagree_bad} compare the first data release imaging products with the re-imaged high-resolution ($<0\farcs.2$) datasets for four sources. 
Figure~\ref{fig:flux_disagree_ok} shows cases where the curve-of-growth profiles become consistent (at the 2 arcsec aperture) or flatten after deeper \textsc{clean}, whereas Fig.~\ref{fig:flux_disagree_bad} presents sources where significant discrepancies remain between the different imaging and flux extraction approaches. These tests suggest that adaptive automasking and deeper \textsc{clean}ing can improve the recovery of extended emission for a subset of high-resolution observations. Approximately 10\% of the full sample show that deeper \textsc{clean} with adaptive automasking improves the imaging. After applying this imaging strategy, fewer than 6\% of sources still show significant discrepancies between the different flux measurement methods.

Given the primary goal of providing a consistent and uniformly processed data release, we retain the fixed aperture measurements as the baseline flux estimates and a uniform \textsc{clean} strategy. We encourage users working with high-resolution datasets to inspect the imaging products carefully and, where necessary, apply imaging strategies optimised for their specific scientific objectives.

\section{Bandwidth distribution and line contamination}

\begin{figure*}
\centering
\includegraphics[width=0.43\textwidth]{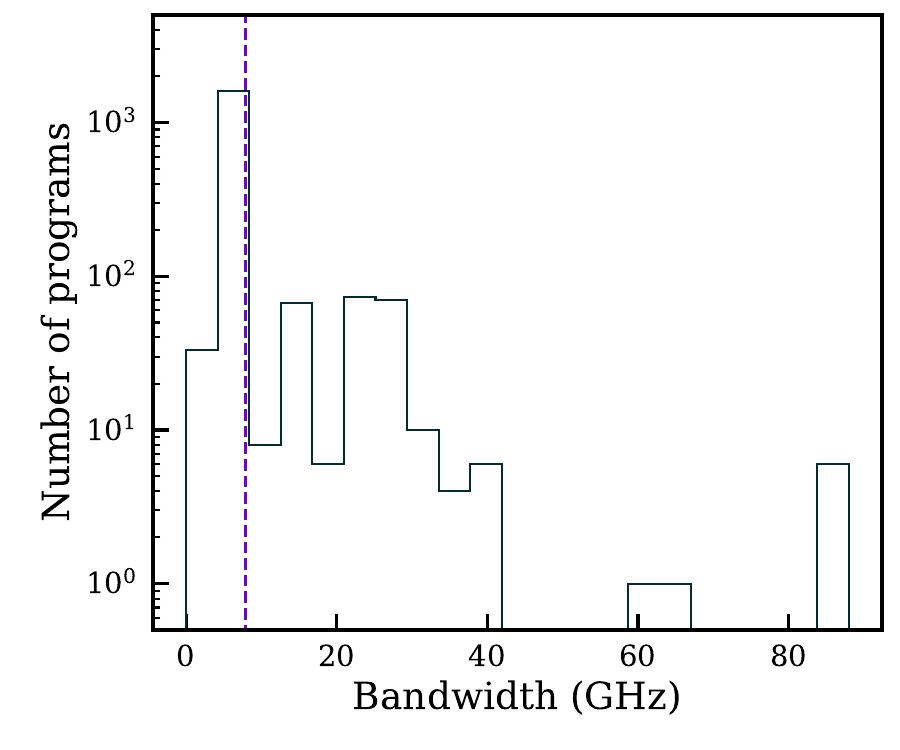}
\caption{Bandwidth distribution. The dashed line shows the median value of the bandwidth distribution at 8 GHz.\label{fig:bandwidth}}
\vspace{1em}
\includegraphics[width=0.98\textwidth]{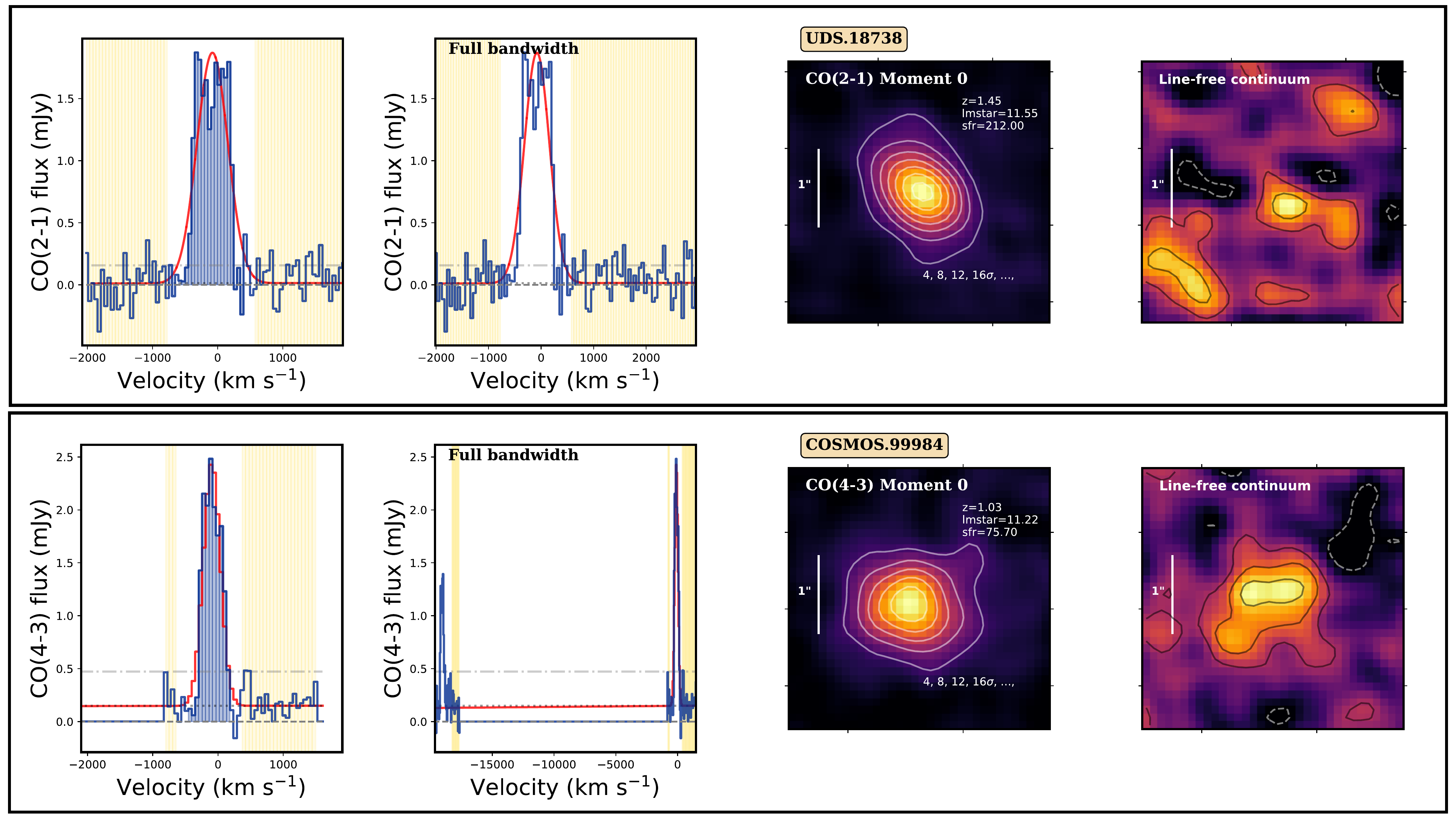}
\caption{Two cases of strong line contamination in the detected continuum. The leftmost panel shows the line profile around the line emission, and the second-left panel shows the entire spectral coverage in the observations. For UDS.18738 (upper panels), the bandwidth is narrower than the typical setup, and hence the measured continuum is dominated by the strong line. For COSMO.99984 (lower panels), there are two strong lines (CO(4-3) and [CI](1-0)) that occupy a large fraction of the bandwidth. The third panel for each galaxy shows the map collapsing the line regions (shaded blue region in the first panel), while the last panel shows the continuum map only taking into account the line-free regions (shaded yellow regions). \label{fig:strongline}.}
\end{figure*}

Figure~\ref{fig:bandwidth} shows the distribution of the bandwidths for individual ALMA programmes.
The median bandwidth is $\sim8$ GHz, which corresponds to the typical bandwidth for a single spectral setup for extragalactic continuum ALMA observations.

The analysis presented in Sect.~\ref{sec:linecontamination} shows that 14\% of the spectroscopically confirmed galaxies have emission-line contributions exceeding 20\% of the measured continuum flux. This relatively high fraction primarily reflects the fact that our analysis is restricted to galaxies with known spectroscopic redshifts. For the majority of sources, however, the expected emission lines occupy only a small fraction of the broad observing bandwidth, thereby limiting their overall contribution to the measured continuum flux. We find that the most strongly contaminated sources are predominantly systems in which two bright transitions fall within the observed bandwidth (e.g. \cofour\ and \ci), or observations with a narrower bandwidth coverage, particularly at higher observing frequencies where transitions such as \cii\ or \oiii\ are observed.
Figure~\ref{fig:strongline} shows two outliers, where the bandwidth is very narrow, and two strong lines lie within the bandwidth coverage.

\section{Description of the released catalogues}
The catalogues released in this version (v1.1) include both ALMA photometry and the derived physical properties running with \eazypy. In Tables ~\ref{tab:catalogue}, we summarise the column descriptions and the corresponding units.

\begin{table*}[t]
\caption{Columns in the photometry catalogues.\label{tab:catalogue}}
\centering
\small
\begin{tabularx}{\textwidth}{l c >{\raggedright\arraybackslash}X}
\hline \hline
\multicolumn{3}{c}{Prior catalogue}  \\ \hline
Column Name           & Unit    & Description                           
                    \\ \hline \hline
projectID             &         & ALMA project ID                                                                                                                 \\
target\_alma          &         & ALMA target names without space; if original ALMA programme has a space in the target name, this column does not include the space \\
small\_mosaic\_idx    &         & mosaic identifier if it was obtained in the mosaic mode                                                                         \\
field                 &         & legacy field names                                                                                                              \\
id\_ecogal            &         & \ecogal\ parent ID                                                                                                                \\
id\_new               &         & \ecogal\ parent full ID including the field name                                                                                  \\
frequency             & GHz     & observed frequency                                                                                                              \\
band                  &         & observed ALMA band                                                                                                              \\
beam\_maj             & arcsec  & synthesised beam major axis                                                                                         \\
frame                 &         & image frame                                                                                                                     \\
RA\_parent            & degree  & prior position from JWST/HST (RA)                                                                                   \\
Dec\_parent           & degree  & prior position from JWST/HST (DEC)                                                                                 \\
RA\_peak\_alma        & degree  & ALMA peak position (RA)                                                                                    \\
Dec\_peak\_alma       & degree  & ALMA peak position (DEC)                                                                                            \\
separation\_prior     & arcsec  & position offset between ALMA peak and the prior position                                                                        \\
flux\_peak            & Jy      & peak ALMA flux                                                                                                           \\
noise                 & Jy      & 1-sigma noise level (after correcting for the primary beam response)                                                            \\
sn                    &         & peak signal-to-noise level                                                                                                      \\
flux\_aper &
  Jy &
  \begin{tabular}[c]{@{}l@{}}flux from the aperture photometry (primary beam corrected), \\ for non-detection, the 3-sigma limit after taking into account the aperture size\end{tabular} \\
eflux\_aper           & Jy      & flux error from the aperture photometry (primary beam corrected); for non-detection, the value is -999                           \\
flux\_peak\_imfit     & Jy      & peak flux from the 2D Gaussian fitting using imfit; if the fit fails, it has a value of -999                                    \\
flux\_imfit           & Jy      & flux from the 2D Gaussian fitting using imfit, if the fit fails, it has a value of -999                                         \\
eflux\_imfit          & Jy      & flux error from the 2D Gaussian fitting using imfit; if the fit fails, it has a value of -999                                   \\
fac\_pbcor            &         & primary beam response at the position of the source; 1=centre                                                                   \\
ecogal\_ver           &         & ecogal data release version                                                                                                     \\
file\_alma            &         & ECOGAL ALMA fits file name (primary beam corrected)                                                                             \\ \hline \hline
\multicolumn{3}{c}{Blind catalogue}  \\ \hline
RA\_bdsf              & degree  & ALMA position (RA) based on PYBDSF fit                                                                                          \\
DEC\_bdsf             & degree  & ALMA position (Decl.) based on PYBDSF fit                                                                                       \\
Total\_flux\_pbcor    & Jy      & Total continuum flux based on PYBDSF fit (corrected for the primary beam response)                                                        \\
E\_Total\_flux\_pbcor & Jy      & Total continuum flux error based on PYBDSF fit (corrected for the primary beam response)                                                  \\
pbfac                 &         & primary beam response at the position of the source; 1=centre                                                                   \\
Peak\_flux            & Jy      & Peak flux from PYBDSF fit (before the primary beam response)                                                                    \\
E\_Peak\_flux         & Jy      & Peak flux error from PYBDSF fit (before the primary beam response)                                                              \\
Maj                   & degree  & the FWHM of the major axis of the source                                                                                        \\
E\_Maj                & degree  & the 1-sigma error on the FWHM of the major axis of the source                                                                   \\
Min                   & degree  & the FWHM of the minor axis of the source                                                                                        \\
E\_Min                & degree  & the 1-sigma error on the FWHM of the minor axis of the source                                                                   \\
PA                    & degree  & the position angle of the major axis of the source measured east of north                                                       \\
E\_PA                 & degree  & the 1-sigma error on the position angle of the major axis of the source                                                         \\
Maj\_img\_plane       & degree  & the FWHM of the major axis of the source in the image plane                                                                     \\
E\_Maj\_img\_plane    & degree  & the 1-sigma error of the major axis of the source in the image plane                                                            \\
Min\_img\_plane       & degree  & the FWHM of the minor axis of the source in the image plane                                                                     \\
E\_Min\_img\_plane    & degree  & the 1-sigma error of the minor axis of the source in the image plane                                                            \\
PA\_img\_plane        & degree  & the position angle in the image plane of the major axis of the source measured east of north                                    \\
E\_PA\_img\_plane     & degree  & the 1-sigma error of the image plane of the major axis of the source measured east of north                                     \\
Isl\_Total\_flux      & Jy      & the total integrated Stokes I flux density of the island in which the source is located (not primary beam corrected)            \\
E\_Isl\_Total\_flux   & Jy      & the 1-sigma error on the total flux density of the island in which the source is located (not primary beam corrected)           \\
Isl\_rms              & Jy/beam & the average background rms value of the island (derived from the rms map)                                                       \\
S\_Code &
   &
  \begin{tabular}[c]{@{}l@{}}a code that defines the source structure; ‘S’ = a single-Gaussian source that is the only source in the island; \\ ‘C’ = a single-Gaussian source in an island with other sources; 'M’ = a multi-Gaussian source\end{tabular} \\
Separation\_bdsf      & arcsec  & offset between the optical counterpart (matched within 0.8 arcsec)   \\ \hline \hline
\multicolumn{3}{c}{Flags}  \\ \hline
flag\_clean      &       & =1: deeper automasking improves the curve-of-growth, and shows a large discrepancy\\
flag\_line      &        & =1: line contribution to the measured continuum flux (flux\_aper) is greater than 20\%.\\
flux\_linefree\_aper      &   Jy     & re-estimated flux excluding the line-emitting channels for line-detected sources with spectroscopic redshift (for a subset of sources discussed in Sect.~\ref{sec:linecontamination}).\\
eflux\_linefree\_aper      &   Jy     & re-estimated flux error for the line-free continuum.\\
\end{tabularx}
\end{table*}

\begin{table*}[t]
\addtocounter{table}{-1}\caption{continued.\label{tab:catalogue2}}
\centering
\small
\begin{tabularx}{\textwidth}{l c >{\raggedright\arraybackslash}X}
\hline \hline
\multicolumn{3}{c}{Ancillary information}  \\ \hline
Column Name           & Unit    & Description                           
                    \\ \hline \hline
id\_3dhst\_v4         &         & 3D-HST(ver4.0) catalogue id when available                                                                  \\
zsp\_best\_avail      &         & best available spectroscopic redshift                                                                                           \\
zsp\_best\_survey     &         & selected spectroscopic survey name                                                                                              \\
z\_ver                &         & DJA MSA version                                                                                                                 \\
objid                 &         & Unique source identifier for the DJA MSA emission line catalogue (unique within the same version)                               \\
srcid                 &         & Source ID from the APT plan                                                                                                     \\
file                  &         & DJA spectrum filename                                                                                                           \\
grating               &         & NIRSpec grating                                                                                                                 \\
file\_phot            &         & Filename of the DJA photometric catalogue \\
id\_phot              &         & ID number in the DJA photometric catalogue \\
valid                 &         & Redshift matches best z from visual inspection                                                                                  \\
z\_phot\_eazy         &         & photometric redshift from eazypy combining 3D-HST+DJA photometry                                                                \\
restU                 & uJy     & flux density of the rest-frame U band from the photoz (eazypy) fit                                                              \\
restU\_err            & uJy     & flux density error of the rest-frame U band from the photoz (eazypy) fit                                                        \\
restV                 & uJy     & flux density of the rest-frame V band from the photoz (eazypy) fit                                                              \\
restV\_err            & uJy     & flux density error of the rest-frame V band from the photoz (eazypy) fit                                                        \\
restJ                 & uJy     & flux density of the rest-frame J band from the photoz (eazypy) fit                                                              \\
restJ\_err            & uJy     & flux density error of the rest-frame J band from the photoz (eazypy) fit                                                        \\
mass\_ez              & Msun    & stellar mass from the eazypy fit; when spec-z is available, redshift is fixed to spec-z                                         \\
sfr\_ez               & Msun/yr & star formation rate from eazypy fit; when spec-z is available, redshift is fixed to spec-z                                      \\
Av\_ez                & mag     & Av from the eazypy fit; when spec-z is available, redshift is fixed to spec-z    

\end{tabularx}
\end{table*}

\end{appendix}
\end{document}